\documentclass[runningheads,a4paper]{llncs}
\usepackage{amssymb}
\usepackage{array}
\usepackage{graphicx}
\usepackage{caption}
\usepackage{subfig}
\usepackage{caption}
\usepackage{float}
\usepackage{wrapfig}
\usepackage{listings}
\usepackage{paralist}
\usepackage{amssymb}
\usepackage{amsmath}
\usepackage{url}
\usepackage{color}
\usepackage{multirow}
\usepackage{hyperref}
\usepackage{enumitem}
\usepackage[nodate]{datetime}

\newcommand{\keywords}[1]{\par\addvspace\baselineskip\noindent\keywordname\enspace\ignorespaces#1}
\newdateformat{mydate}{\monthname[\THEMONTH] \THEYEAR}
\nonstopmode

\begin{document}
\mainmatter

\title{Dimming the Internet}
\subtitle{Detecting Throttling as a Mechanism of Censorship in Iran}
\author{Collin Anderson\thanks{This project received grant funding from the Center for Global Communication Studies at the University of Pennsylvania's Annenberg School for Communcation and Google Research.}}
\date{\mydate\today}
\titlerunning{Dimming the Internet}
\authorrunning{Collin Anderson}
\institute{
  \texttt{\href{mailto:collin@averysmallbird.com}{collin@averysmallbird.com}}
}

\maketitle
\begin{abstract} 
In the days immediately following the contested June 2009 Presidential election, Iranians attempting to reach news content and social media platforms were subject to unprecedented levels of the degradation, blocking and jamming of communications channels. Rather than shut down networks, which would draw attention and controversy, the government was rumored to have slowed connection speeds to rates that would render the Internet nearly unusable, especially for the consumption and distribution of multimedia content. Since, political upheavals elsewhere have been associated with headlines such as ``High usage slows down Internet in Bahrain'' and ``Syrian Internet slows during Friday protests once again,'' with further rumors linking poor connectivity with political instability in Myanmar and Tibet. For governments threatened by public expression, the throttling of Internet connectivity appears to be an increasingly preferred and less detectable method of stifling the free flow of information. In order to assess this perceived trend and begin to create systems of accountability and transparency on such practices, we attempt to outline an initial strategy for utilizing a ubiquitious set of network measurements as a monitoring service, then apply such methodology to shed light on the recent history of censorship in Iran.
\end{abstract}

\section{Introduction}
\label{sec:introduction}
{
\begin{center}
\textit{"Prison is like, there's no bandwidth."}{ - Eric Schmidt}\footnote{Remarks before Mobile World Congress. February 28, 2012}
\end{center}
}

The primary purpose of this paper is to assess the validity of claims that the international connectivity of information networks used by the Iranian public has been subject to substantial throttling based on a historical and correlated set of open measurements of network performance. We attempt to determine whether this pattern is the result of administrative policy, as opposed to the service variations that naturally occur on a network, particularly one subject to the deleterious effects of trade restrictions, economic instability, sabotage and other outcomes of international politics. Overall, we outline our initial findings in order to provoke broader discussion on what we perceive to be the growing trend of network performance degradation as a means of stifling the free flow of information, and solicit feedback on our claims in order to create universally applicable structures of accountability. Furthermore, to the fullest extent possible, we focus our assessment on that which is quantitatively measurable, and limit attempts to augur the political or social aspects of the matters at hand. As in any other closed decision-making system, a wealth of rumors dominate the current perception of the actions taken by the government and telecommunications companies. Where these rumors are mentioned, they are discussed in order to test validity, and not cited as evidence.

This paper is not intended to be comprehensive, and we err on the side of brevity where possible. Toward these ends, our contributions are threefold:

\begin{enumerate}[leftmargin=1cm,rightmargin=.5cm]
    \item outline a methodology for the detection of the disruption of network performance and infer purposeful intent based on indicators, differentiated from normal network failures;
    \item begin to identify potential periods of throttling, based on available historical data;
    \item attempt to enumerate those institutions that are not subject to interference.
\end{enumerate}

The experiments described herein are motivated toward collecting initial, open-ended data on an opaque phenomenon; where possible our results and code are publicly available for outside investigation at: 

\begin{quote}
    \url{http://github.com/collina/Throttling}
\end{quote}

The remainder of this paper is structured as follows. In the following section, we identify the infrastructural properties of networks relevant to our line of inquiry that enable states and intermediaries to control access to content and service performance. We describe the dataset core to our investigation in Section \ref{sec:setup} and Section \ref{sec:mathishard} describes a mixed methods approach used to interpret measurements and extract broader information on the nature of the network. Finally, these techniques are applied in Section \ref{sec:findings} toward identifing periods of significant interest in the connectivity of Iran-originating users. We conclude by enumerating the outstanding questions and future research directions.

\section{Domestic Network Structure Considerations}
\label{sec:infrastructure_considerations}

After the declaration that the incumbent president, Mahmoud Ahmadinejad, had won a majority in the first round of voting, supporters of reformist candidates rallied against what was perceived to be election fixing in order to preserve the status quo of the power structure of the state. Already well-acquainted with bypassing Internet filtering using circumvention and privacy tools, such as VPNs and Tor, government blocks on YouTube and Facebook were minor impediments for activists to share videos and news in support of their cause. Unused to large-scale challenges against the legitimacy and integrity of the system, the government appears to have responded by ordering the shutdown of mobile phone services, increased filtering of social media sites and the disruption of Internet access \cite{bailey2011censorship}.

\begin{figure}[h]
 \centering
  \includegraphics[width=\textwidth]{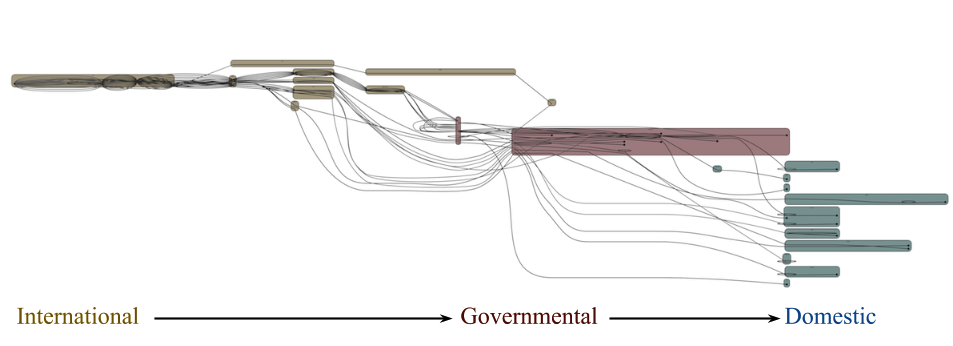}
 \caption{International Pathways to Iranian Hosts (Traceroute)}
 \label{fig:pathways}
 \end{figure}

Iran's telecommunications infrastructure and service market differs substantially from the regulatory environment of broadcast television, wherein the state maintains an absolute monopoly on authorized transmissions \cite{imp:control}. The current  on network ownership was initially shaped by the Supreme Council of the Cultural Revolution in 2001 under the directives ``“Overall Policies on Computer-based Information-providing Networks'' and ``Regulations and Conditions Related to Computerized Information Networks'' \cite{cso_trc,ihrdc:ctrl}. Internet Service Providers (ISPs), which offer last-mile network connectivity, are privatized, but subject to strict licensing requirements and communications laws that hold companies liable for the activities of their customers.\footnote{While ISPs are the public face of Internet access to the Iranian public, these companies are only one component of a broader domestic telecommunications infrastructure responsible for delivering domestic and international traffic. ISPs interconnect with each other, known as \textit{peering}, to provide accessibility to hosts within their network and share routes for traffic between others. Not all ISPs are consumer-facing, with some infrastructure companies acting as dedicated Internet exchange points (IXPs) between networks.} In addition to administrative requirements for filtering according to a nebulous and growing definition of subjects deemed `criminal,' ISPs are forced to purchase their upstream connectivity from government-controlled international gateways, such as the state-owned telecommunications monopoly, the Telecommunication Company of Iran (TCI), which appear to implement an auxiliary, and often more sophisticated, censorship regime on traffic in transit across its network. Even amongst privately-owned networks, perspective entrepreneurs appear to be encouraged or coerced into ownership consortiums with the ever present set of Bonyads, charitable trusts often connected with the Iranian ideological establishment \cite{turkcell}. While a substantial amount has been written on controls imposed on content, the salient principles to our domain of research are the obligations of independent providers to the state's administrative orders and the infrastructural centrality of two entities in Iran's network, the TCI's subsidiary, the Information Technology Company (ITC, AS12880), and the Research Center of Theoretical Physics \& Mathematics (IPM, AS6736). As a result, all traffic to foreign-based hosts, and likely a majority of connections internally, route through entities with either direct or informal relationships to the government, as demonstrated in Figure \ref{fig:pathways}.
 
The centralization of international communications gateways and domestic peering (the linkages between networks) enables anticompetitive and potentially undemocratic practices that would be more administratively difficult and economically expensive in an open and multi-stakeholder telecommunications market. This design is not specific to Iran alone nor is it an indicator of a government's desire to control citizen access. Centralization often resembles a commonplace model of public-sector infrastructure development or state revenue generation from telecommunications surcharges \cite{roberts2011mapping}. With the introduction of the Internet, regulators reflexively extended their mandate to include online communications, as the most common forms of physical connectivity often utilize telephony networks, as well as bringing competitive services such as voice-over-IP. When the network monitoring company Renesys addressed this topic in response to the Internet shutdowns of Egypt and Syria, they framed the dangers and fragility of the centralization of gateways as ``the number of phone calls (or legal writs, or infrastructure attacks) that would have to be performed in order to decouple the domestic Internet from the global Internet,'' naming 61 countries at `severe risk' for disconnection \cite{renesys:couldit}.

Although disconnection, failure and filtering are more perceivable forms of disruption, the same principles of risk and exposure apply to the degradation of connections. Throttling is not on its own a form of censorship or intent to stifle expression. In the context of Iran, a scarcity of available bandwidth and inadequate infrastructure has created a demonstrable need for limiting resource-intensive services and prioritizing real-time communications traffic, especially in rural markets \cite{hassani2010qos}. In other cases, often under the terms ``quality of service'' or ``traffic shaping,'' throttling is a means of providing higher performance to less bandwidth-intensive applications, through initially provisioning of faster speeds to a connection that is then slowed after a threshold is reached. These practices have spurred heated debate between civil society and telecommunications providers in the United States and Europe within the framework of ``network neutrality,'' pitting the core principle that communications on the Internet be treated equally against claims by companies that the bandwidth demands of online services exceeds current availability. However, the allegations of throttling we attempt to address differ substantially from these debates in scope and execution, and fit into a history of interference with the free flow of information, offline and online, and recurrent security intrusions on the end-to-end privacy of the communication of users, often originating from government-affiliated actors \cite{foxitdiginotar}.

\section{Setup}
\label{sec:setup}

Degradations in network performance and content accessibility can be the product of a number of phenomenon and externalities, localized to one point or commonly experienced across a wider range of Internet users. In order to accurately and definitively measure broadly-targeted degradation, such as throttling, it is necessary to obtain data from a diversity of hosts, in terms of connection type, physical location, time of usage and nature of usage.

A number of tools have been developed to actively probe qualities of infrastructure directly relevant to throttling and disruption \cite{dischinger2010glasnost,filasto2012ooni,kreibich2010netalyzr}. These techniques compare whether traffic flows of differing types sent at an identical rates are received differently, thereby comparing against an established baseline to give a clear indication of potential discrimination. Where such data has been core to network neutrality debates, for countries such as Iran, government opacity on broadband deployment means that domestic civil society and private parties have had few opportunities to embrace quantitative data to push for policy changes. Our research interest biases observations that are ubiquitous, recurrent and not necessitating the intervention of users, in order to plot historical trends and account for localized aberrations, even if at the cost of precision or confidence. As a result, to meet our operational needs, we resort to the use of measurements that, while not specifically designed to detect throttling, broadly assess relevant characteristics of the network in a manner that may indicate changes in the performance and nature of the host's connectivity.

In order to collect a statistically significant set of network performance datapoints, we utilize the data collected by client-initiated measurements of the Network Diagnostic Tool (NDT) hosted by Measurement Lab (M-Lab) \cite{dovrolis2010measurement}, which contains both a client-to-server (C2S) and server-to-client (S2C) component. The throughput test consists of a simple ten second transfer of data sent as fast as possible through a newly opened connection from a M-Lab server to a NDT client. In addition to measuring the throughput rate, the NDT test also enables the collection of diagnostic data that can assess factors such as latency, packet loss, congestion, out-of-order delivery, network path and bottlenecks on the end-to-end connectivity between client to server.

 \begin{figure}[h!]
 \centering
  \includegraphics[width=\textwidth]{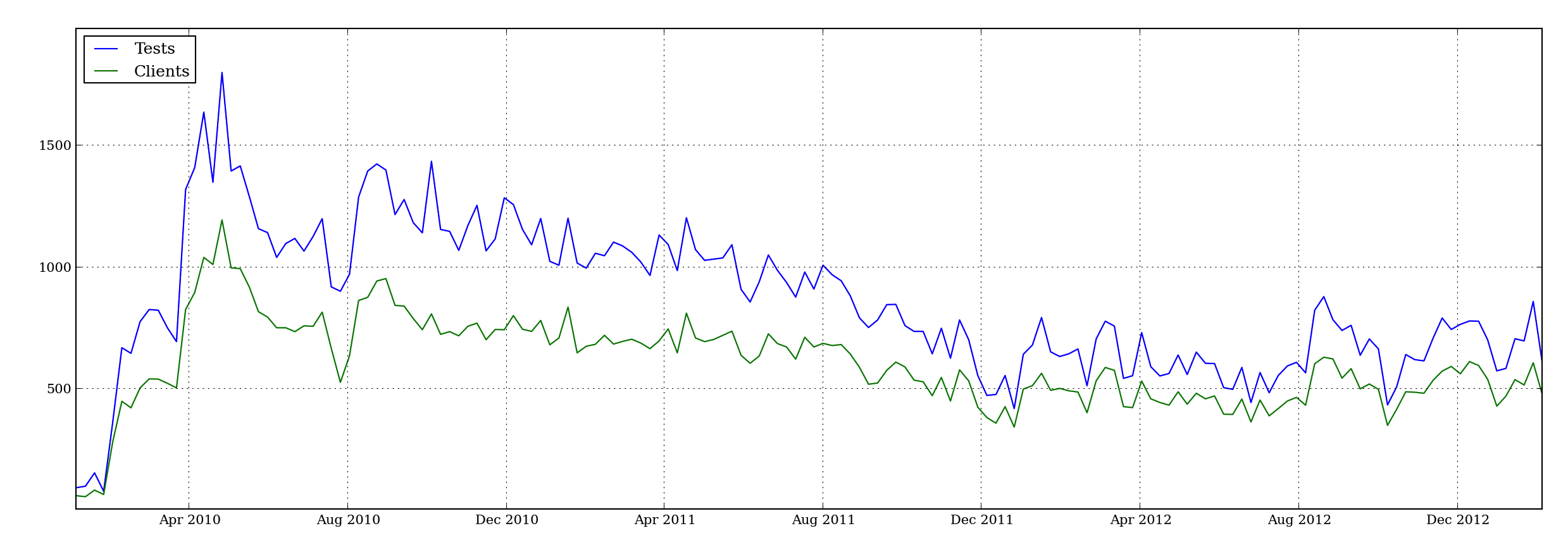}
 \caption{Unique Clients and Tests from Iran (Weekly), Jan 2010 - Jan 2013}
 \label{fig:clientstests}
 \end{figure}

M-Lab's added value is both methodological and institutional. Founded and administered by a consortium of non-governmental organizations, private companies and academic institutions, including New America Foundation's Open Technology Institute, PlanetLab, and Google, M-Lab is an open data platform focused on the collection of network measurements related to real-world broadband connectivity, and is not in itself a human rights or political cause. M-Lab's data is both non-partisan and widely-accepted, having been used by telecommunications regulators and development agencies in Austria, Cyprus, the European Commission, Greece and the United States. With the inclusion of NDT as a connectivity diagnostic test in version 2.0 of the Bittorrent file-sharing client $\mu$Torrent, M-Lab gained a significant increase in NDT measurements and a wider audience of users\cite{blog:utorrent}. Since 2009, M-Lab has collected over 725 Terabytes of data based on around 200,000 tests conducted per day. More relevant to our purposes, in the month of January 2013, M-Lab collected 2,925 tests from 2,158 clients from Iran that we consider valid under the definitions and parameters described in Section \ref{sec:mathishard}. Figure \ref{fig:clientstests} demonstrates the rapid growth of network measurements originating from Iran after the inclusion of NDT in $\mu$Torrent, with unique clients representing the number of IP addresses seen and tests describing the total number of tests. While M-Lab contains measurements from 2009, we consider data beginning in 2010 to take advantage of the critical mass of clients that resulted from the inclusion.

During the process of preparing and running our experiments, we took special care to not violate any laws or, considering the diminishing opportunities for international collaboration, expose individuals within Iran to potential harm. All our experiments were in accordance to any applicable terms of service and within reasonable considerations of network usage, taking care to not engage in behavior that would be considered intrusive. The contribution of measurements of a network to NDT or M-Lab does not denote political activities on the part of the user, particularly as the use of Bittorrent file-sharing has broad appeal and apolitical implications. Moreover, use of Bittorrent or NDT do not appear to violate Iranian law, especially given the copyright and intellectual property framework of Iran. All data originating from M-Lab is openly available to the public, and collected in a manner that does not reveal potentially identifiable user information, outside of the client's IP address. 

\section{Parameters \& Calculations}\label{sec:mathishard}

NDT provides a multiplicity of metrics and diagnostic information that describe the test session. As outsiders attempting to assess the behavior and actions of network intermediaries, we are relegated to beginning from assumptions and hypotheses founded on general principles of networking and how others accomplish throttling. Additionally, we are limited in the lessons to be gained from the corpus of research on assessing violations of network neutrality, as we struggle to establish control measurements and perform the active probing necessary for proper comparative analysis. Most methods for measuring broad or application-specific throttling assume that connections are initially allocated a higher level of throughput, which is then throttled or terminated upon identification by a network intermediary or exceeding a bandwidth quota. Therefore, our primary form of detection of abnormal network conditions is comparative assessment of selected indicators based on historical trends and incongruities between subgroups of clients. In order to allow this comparative assessment through baselines that serve as controls, we  define a set of measurements and aggregate clients in a consistent and non-abitrary manner.

For the introductory purposes of this study, and based on cursory analysis of existing M-Lab data which was generated during suspected throttling events, we substantially rely on NDT's measurements of round trip time, packet loss, throughput and network-limited time ratio as potential identicators of network disruption.\footnote{For the purpose of space we have abbrievated the NDT-recorded variables of SndLimTimeRwin, SndLimTimeCwnd, SndLimTimeSnd}

\begin{description}
\item[Round Trip Time (RTT)] ($MinRTT, MaxRTT, \frac{SumRTT}{CountRTT}$)\\ 
The time taken for the round trip of traffic between the server to client, computed as the difference between the time a packet is sent and the time an acknowledgement is received, also known as latency. There are a diversity of causes for latency, including `insertion latency' (the speed of the network link), the physical distance of the path taken, `queue latency' (time spent in the buffer of network routers), and `application latency.' The last of these components, application latency, is accounted for within the NDT test under the metrics of \textit{Receiver and Sender-limited time windows}.  We are interested in latency that occurs due to network properties, primarily the queue, insertion and path latencies.  Since the time taken should not change dramatically, fluctuations indicate a meaningful change in connectivity, such as a network outage that increases latency due to traffic to taking a longer route and telecommunications equipment having taken on additional load.  NDT also provides different approaches to this measurement, which allow alternative perspectives on the network. The minimum round trip time record (MinRTT) mostly occurs before the network reached a point of congestion, and therefore is generally not thought to be indicative of real performance. Alternatively, the average of RTTs, through the division of the sum of all round trip times by the number of trips may more closely approximate latency, but is also vulnerable to outlier values.
\cite{findingthelatency}

\item[Packet Loss] ($\frac{CongSignals}{SegsOut}, \frac{SegsRetrans}{DataSegsOut}$)\\ 
The transmission of traffic across a route is not guaranteed to be reliable, and network systems are designed to cope with and avoid failure. These mechanisms include maintaining an internal timer that will give up on traffic the system has sent, alerts from the network that congestion is occurring, and notification from the other end that data is missing. Packet loss for our purposes is defined as the the number of transmission failures that occurred, due to all forms of congestion signals recorded under NDT's test, including \textit{fast retransmit}, \textit{explicit congestion notifications} and \textit{timeouts}. In order to address this relative to the amount of extensiveness of the test, packet loss is measured as a probability against the number of packets sent.

\item[Network-Limited Time Ratio] ($\frac{TimeCwnd}{TimeRwin + TimeCwnd + TimeSnd}$)\\ 
The network stack of operating systems maintain internal windows of the amount of traffic that has been sent and not acknowledged by the other party. This enables the system to avoid over-saturating a network with traffic and to detect when a failure has occurred in communications.  The NDT test attempts to send enough traffic to create congestion on the network, where the transmitting end of traffic exceeds this window of unacknowledged traffic and is waiting for clearance from the other side to continue sending. The Network-Limited Time Ratio is calculated as the percentage of the time of the test spent in a `Congestion Limited' state, where sending of traffic by the client or server was limited due to the congestion window.

\item[Network Throughput] ($\frac{HCThruOctetsAcked*8}{TimeRwin + TimeCwnd + TimeSnd}$)\\
The NDT test attempts to send as much data as quickly as possible between the client and an M-Lab server for a discrete amount of time in order to stress the capacity of the network link.  For the upload performance, this is calculated based of the amount of data received from the client, and with the download being the number of sent packets that were acknowledged as received. The throughput rate is then calculated against the time that the test lasted. 
\end{description}

Measurements of connection properties allow for the indirect inference of broader network conditions, particularly when applied in a comparative fashion. For example, others have noted the varying degrees of correlation between round trip time and the total load presented on a network\cite{biaz2003round}. While these studies hold a higher correlation on a slow link than on a fast one, the former of which appears to more accurately describe the domestic connectivity of Iran. In this scenario, it may be possible to identify periods where an increase in load has created a bottleneck in the network or a build-up in the traffic queue on network devices. Therefore, we assume that the decrease of throughput, or increase of loss, will be associated with an increase in latency where traffic congestion is occurring. Additionally, while most inter-network routing protocols will attempt to route traffic over the best path to a destination, which can differ based on load balancing and service agreements, a change in round trip time may indicate a change in the path traversed by the data sent from the client to M-Lab. Due to the variability in network conditions that can affect latency, we use two measurements: the minimum RTT recorded in the session and the uniform average over the entire test.

\subsection{Aggregation and Comparative Methods}

In order to assess the general performance of the domestic network, client measurements are aggreggated across higher-level groupings and evaluated based on their median value. Where multiple tests are performed by a client during an evaluation period, the most performant measurement is used in order to mitigate potential bias in samples. Through giving preference to faster measurements, we intentionally bias our pool of data against our hypothesis and assume that less favorable numbers are aberrations. Furthermore, we prefer calculations that accomodate for outlier values, such as false positives that occur in geolocation services when foreign-hosted networks are registered to domestic entities. Finally, we assume the natural shifts in consumer behavior or infrastructure development that may affect measurements, such as the adoption of mobile broadband, are gradual and upward trends. In practice this assumption appears to not only hold, but we are led to question whether the availability of high-speed connectivity has declined, due to administrative limitations imposed on consumer providers and delays in the development of mobile data licensing.

For our purposes, we discretely aggregate measurements across three dimensions related to the character of the tests or location of the client,

\begin{description}
\item[National:] Measurements are grouped on a country level. Aggregation for large geographic areas or service providers may represent a diverse strata of connectivity types, such as ADSL, dialup, WiMAX and fibre. 

\item[Internet Service Providers and Address Prefixes:]  We use the Autonomous System Number (ASN) as a proxy for ISPs. Within a research methodology, aggregation based on the ASN provides a larger pool of clients at the cost of being less granular than address prefixes.  Autonomous Systems are generally defined as a set of routers under a single technical administration, which keep an understanding of global network routes to direct traffic and announce their ownership of blocks of IP addresses (address prefixes). There is a limited pool of available numbers for the labeling of Autonomous Systems, and not every network has the need to advertises its own set of routing policies, particularly where directly connected networks have the same upstream connection. Therefore, a large ISP, such as Afranet, will generally maintain one or more Autonomous Systems, bearing the responsibility to maintain announcements and peering, for the connectivity leased to other smaller ISPs, government agencies, educational institutions or commercial organizations. Additionally, since traffic paths across networks are constructed using centrally allocated components of ASNs and address prefixes, both are registered and externally queriable. 

\item[Control Groups:] We attempt to identify logical, coherent groups of networks and clients based on common characteristics, such as the nature of the end user or performance. Control group measurements differ from service providers because they are defined as narrowly as the data allows, rather than existing segmentations. Such groups, particularly when defined by network degradation, are possibly deterministic, however, these often produce surprising and mixed results, as described in Section \ref{sec:findings}. Furthermore, the grouping of networks or entities, such as state agencies or educational institutions, during one incident is testable under other circumstances and may serve as a control for future monitoring. 

\end{description}
 
Finally, detection of significant events generally follows one of two themes. These mechanisms are designed to highlight precipitous changes of service quality as a warning system to flag potential events, however, they are not the holistic determinant of interest.

\begin{figure}
 \centering
  \includegraphics[width=\textwidth]{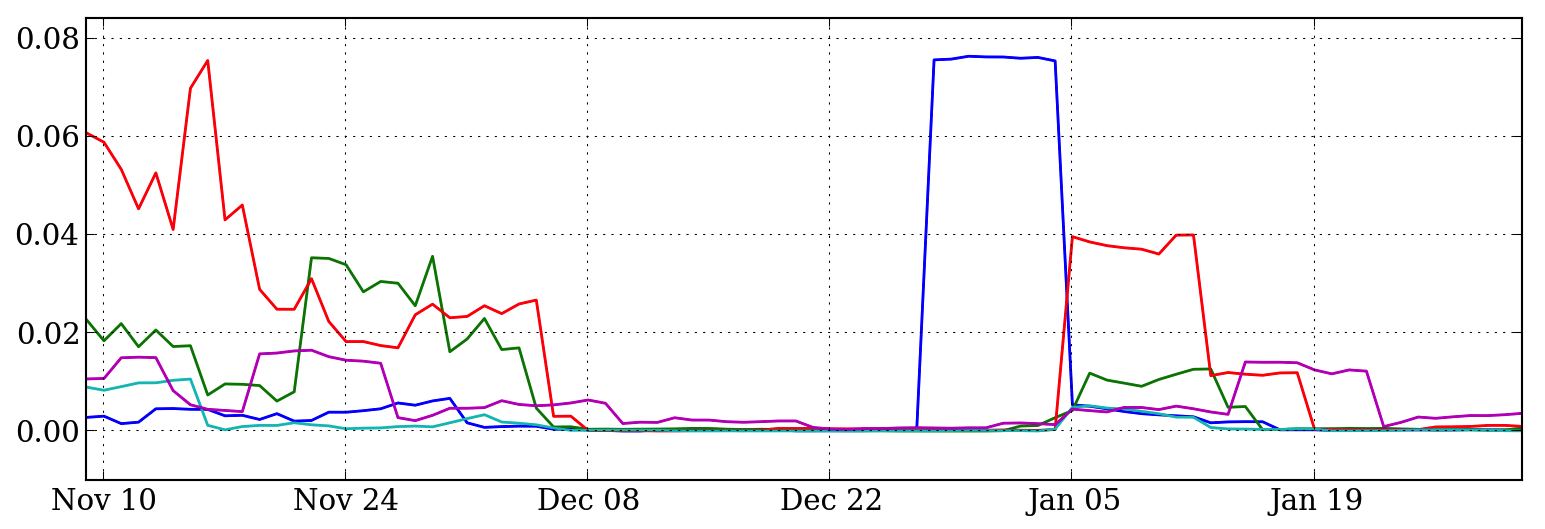}
 \caption{Throughput Variance on Daily Medians per ASN, November 2011 - January 2012}
 \label{fig:nov11:thuvar}
 \end{figure}

\begin{description}
\item[Threshold:] A maximum and minimum threshold of reasonable values are established based on previous trends. Since our dataset is frequently limited to a small amount of clients that are subject to varying network conditions and unrelated externalities, producing wide fluctuations in measurements, a Poisson distribution is established based on a rolling average. Detection of an abnormal event occurs when the trend breaches these bounds \cite{danezis2011anomaly}.

\item[Variance:] Internet Service Providers offer a diversity of products with varying levels of performance across different markets, leading to variations in qualities, such as connection speed and reliability. This technological and commercial variation is exacerbated by informal differences, such the scrutiny placed on the documentation necessary to obtain faster broadband implementation of administrative orders (elaborated in Section \ref{sec:controlgroups}) and ability to acquire network infrastructure, despite scarcity created by sanctions and exchange rates. While these differences may change across time, there should be a consistent trend of diversity within a free market. We evaluate the variation that occur in service quality amongst our subgroups such as ISPs or prefixes. Therefore we presume that when external limitations are not imposed, variation will be high, while the contrapositive holds that at a time of control, the variation will be low. We use the classic variance of the average of the squared deviations from the mean to accomodate proportional change.

Analysis based on the variance of performance measurements day over day across short historical periods can be applied to a single network or client as between ISPs. We would anticipate that if an administrative ceiling were imposed on throughput speed, particularly at a limit below the potential capacity of the network, the variation would near zero, as the tests would cap out at the maximum available bandwidth. We would also expect that trend line of such an incident to follow a peak and valley model, where a sudden decrease or increase leads to a spike in variance when the limit is imposed or lifted, with low variance during the throttling. Figure \ref{fig:nov11:thuvar}, demonstrates that in practice this hypothesis holds mixed results. While peaks do occur, the variance, particularly the relative variance, remains high.

\end{description}

As these mechanisms serve as a warning system to direct further investigation, rather than being a sole determinant of interest, we are less concerned about its robustness. Additionally, for the purposes of identifications and coding of events, we generally call attention to extremely abnormal values, relative to the normal trends. Once a significant change in service is identified, it is subject to correlation against other metrics. More constrained boundaries simply lead to more research cost and false positives. We assume throttling designed to stifle expression or access is not a subtle event.

\subsection{Limitations}

While the integration of NDT into $\mu$Torrent has enabled the massive proliferation of points of observation across a variety of geographic locations and network conditions, we remain relegated to user-contributed data collection based on a limited set of volunteers. Changes in connectivity or quality of service cannot directly be infered as an administratively-imposed censorship event. Additionally, the Network Diagnostic Test's data collection does not occur in isolation of other externalities that are likely to affect performance. Cross traffic, local and upstream network activities from other applications or users, independent of the test can bias results through introducing additional latency or failures as the test mechanism competes for bandwidth and computing resources. As an end user is positioned at the border of the network, they cannot independently account for the conditions outside of their control that may impact the results of this test. Therefore, like others before, we attempt to mitigate and account for externalities based on interpretations derived from observable data and general networking principles \cite{salamatian2003cross,weinsberg2011inferring}.

 \begin{figure}
 \centering
 \includegraphics[width=\textwidth]{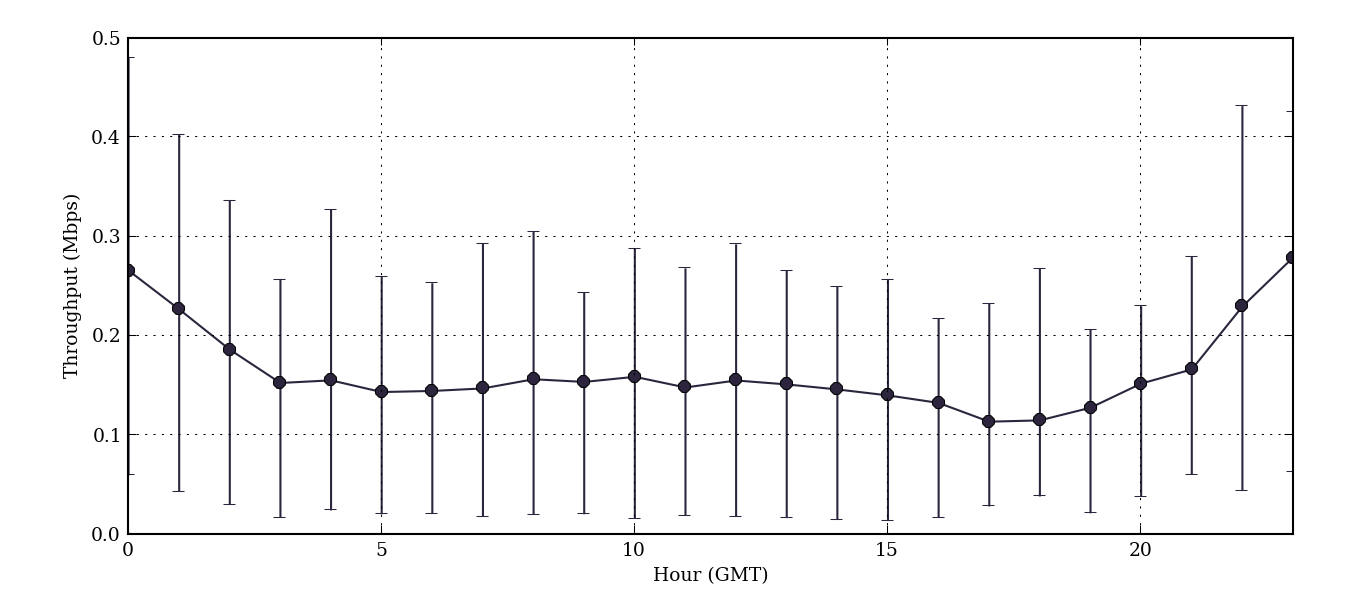}
 \caption{Diurnal Patterns of Throughput Measurements, Iran, Jan - Mar 2013}
 \label{figure:mathishard:diurnal}
 \end{figure}

The extent of externalities often vary based on diurnal patterns of use and in response to specific incidents. Using the performance metric of throughput, Figure \ref{figure:mathishard:diurnal} demonstrates that observed network speed is higher in the early hours of the morning, Iran time (GMT +4:30), than during the day. These trends are mirrored in the performance across the week, and within other measurements, such as packet loss. It would appear that Iran's network does not handle the additional load of office hours and evening use gracefully.

M-Lab-based services attempt to perform testing under favorable conditions by selecting the measurement server by geographic proximity. These servers are physically located across the world, including in Australia, Austria, Czech Republic, France, Germany, Greece, Japan, Ireland, Italy, the Netherlands, New Zealand, Norway, Spain, Slovenia, Sweden, Taiwan, the United Kingdom and the United States. Although one of these M-Lab hosts countries directly peers with Iran,\footnote{Based on Hurricane Electric's public data of the peering of AS12880, of these countries, Iran is only directly connected with Italy} the selection mechanism predominantly directed NDT clients to Greece (of the 2921 tests in January 2013, 50.39\%), United States (22.08\%), United Kingdom (16.40\%) and France (9.28\%), with the remaining three countries constituting 1.85\% of tests. It may be possible that if there were M-Lab servers in Turkey or Azerbajian, NDT tests would result in higher throughput measurements, however, we derive our conclusions from relative changes, rather than absolute numbers. Addditionally, users that are connected to anti-filtering tools during the test are likely to be recorded by M-Lab as originating from the country that the tool routes its traffic through, and therefore not included in our sample. 

We also constrain our expectations on the types of throttling or disruption that NDT will detect. As a diagnostic of direct connectivity between hosts, the test is based on traffic patterns that likely have yet to inspire scrutiny from intermediaries. Other protocols, such as those employed in VPN tunnels, HTTP proxies, Tor, voice-over-IP and streaming media, have at varying times been claimed to be subject to targeted interference, based on port or deep packet inspection. NDT's measurement methodology is unlikely to detect more sophisticated discimination against specific forms or destination of traffic. 

Lastly, we assume that intermediaries have not sought to interfere or game the data collection of M-Lab through artifically biasing measurements from hosts. 

\section{Findings}\label{sec:findings}

Using median country-level throughput, evaluated based on the most performant measurement per client per day, we find two significant and extended periods of potential throttling within our dataset, occuring \textit{November 30 2011 - August 15 2012} (a 77\% decrease in download throughput) and \textit{October 4 - November 22 2012} (a 69\% decrease). We identify an additional eight to nine short-term instances where the throughput or variance between providers underwent a precipituous change, triggering the attention of detection mechanisms. These events are correlated with a reduction of service quality across all networks, often more significantly impacting home consumers than commercial institutions. In most cases, these changes mirror more overt increases of interference of communications channnels. Lastly, within available indicators or traffic routes, we do not find evidence that these fluctuations are the result of externalities, such as changes to international connectivity or domestic network use.

\subsection{Periods of Significant Interest}\label{sec:throttling_events}

\begin{figure}
\centering
\subfloat[Throughput Trends]{
\includegraphics[width=\textwidth]{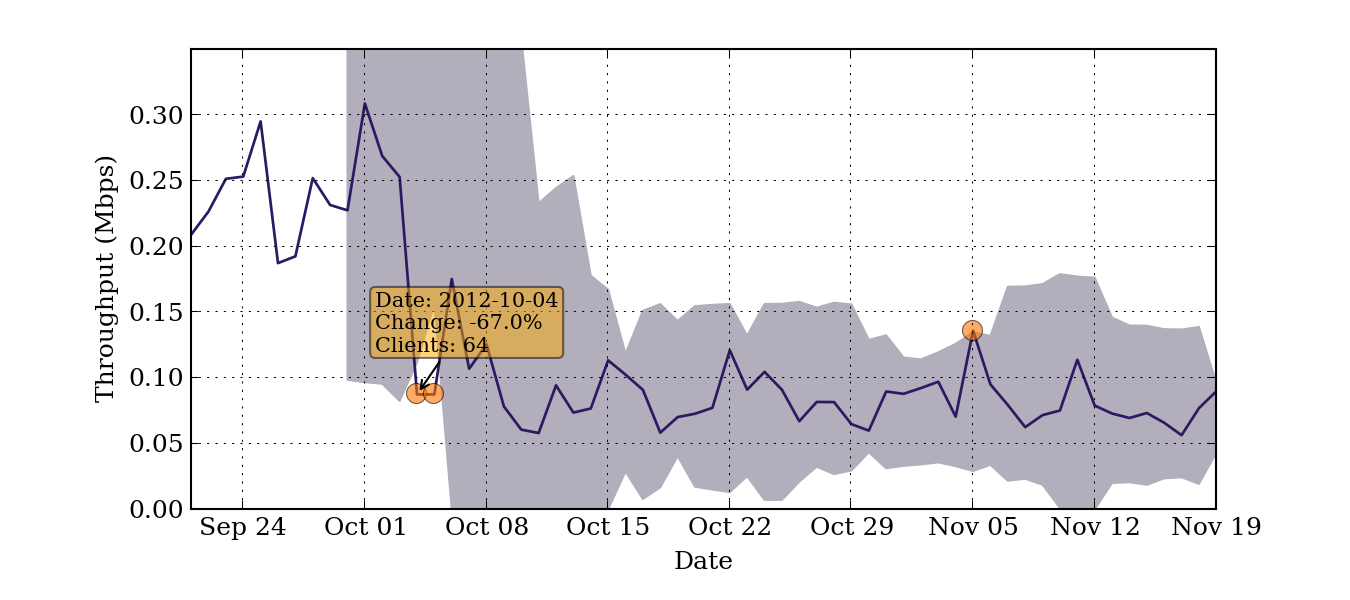}
\label{findings:throttling_events:oct_2012:throughput}
}\hfill
\subfloat[Average and Minimum Round Trip Time]{\includegraphics[width=\textwidth]{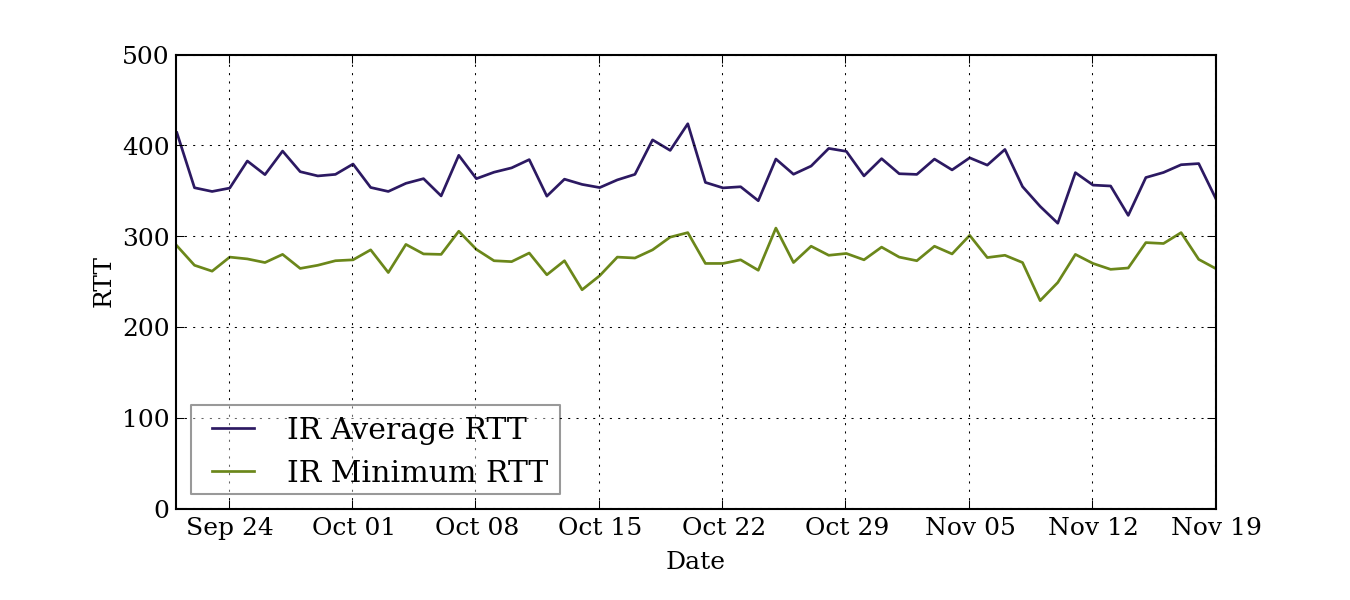}
\label{findings:throttling_events:oct_2012:rtt}
}\hfill
    \subfloat[Throughput Histogram]{
        \includegraphics[width=\textwidth]{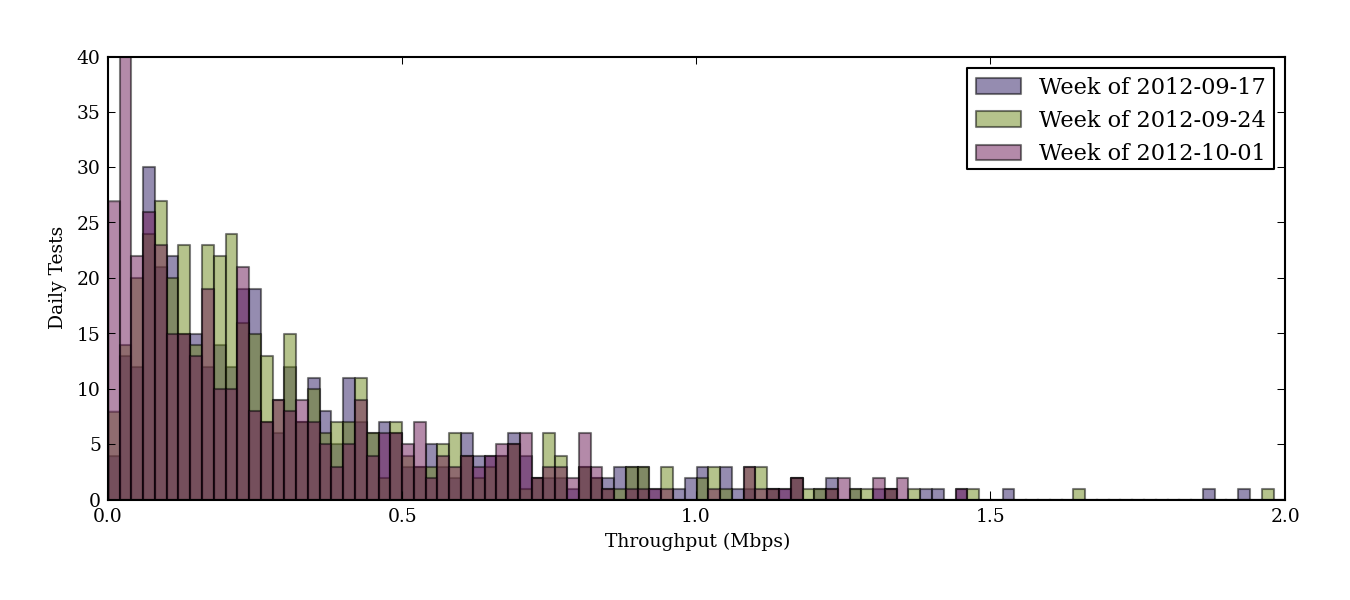}\hfill
    }
    \caption{Suspected Throttling Event, Sep 2012 - Nov 2012}
    \label{findings:throttling_events:oct_2012}
\end{figure}

\begin{figure}
    \centering
    \subfloat[Throughput]{
        \includegraphics[width=\textwidth]{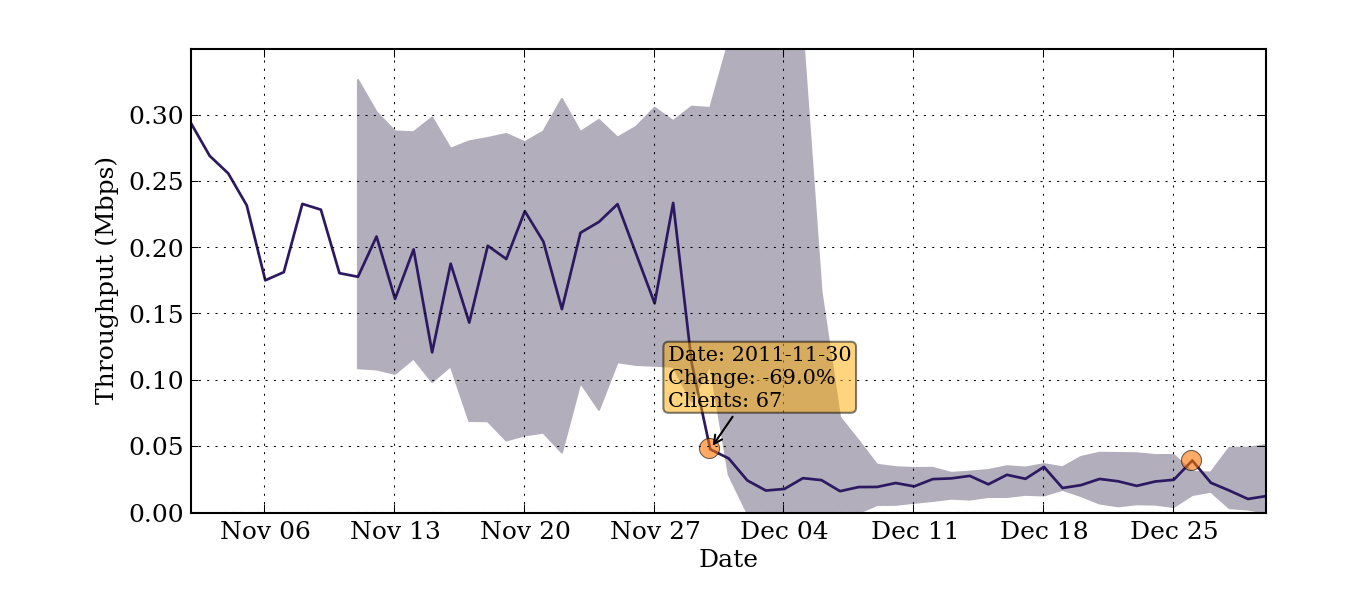}\hfill
    }\hfill
    \subfloat[Average and Minimum Round Trip Time]{
      \includegraphics[width=\textwidth]{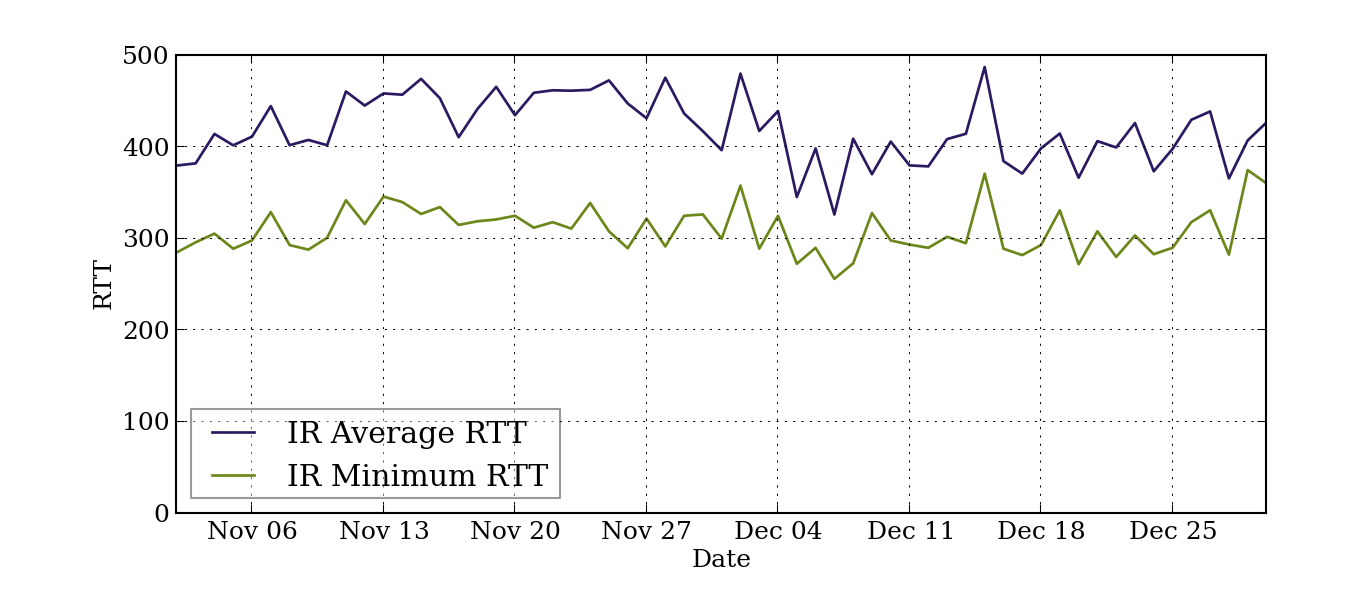}
    \label{findings:throttling_events:nov_2011:rtt}
    }\hfill
    \subfloat[Variance]{
        \includegraphics[width=.5\textwidth]{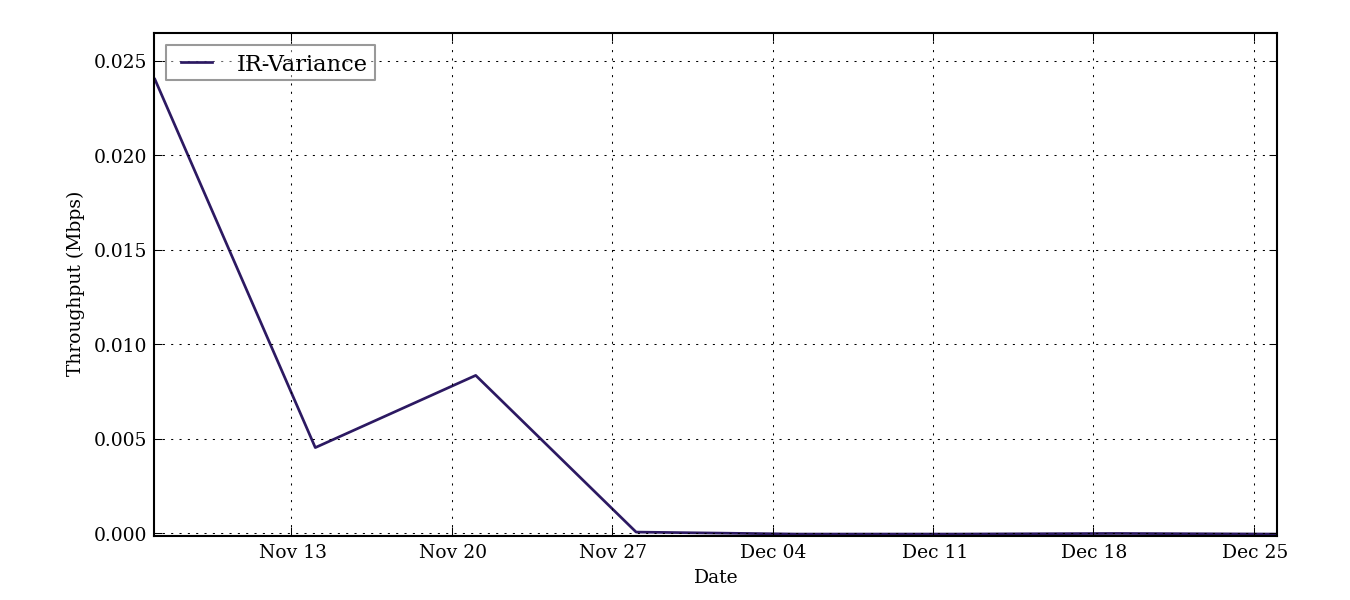}\hfill
    }
    \subfloat[Network Congestion]{
        \includegraphics[width=.5\textwidth]{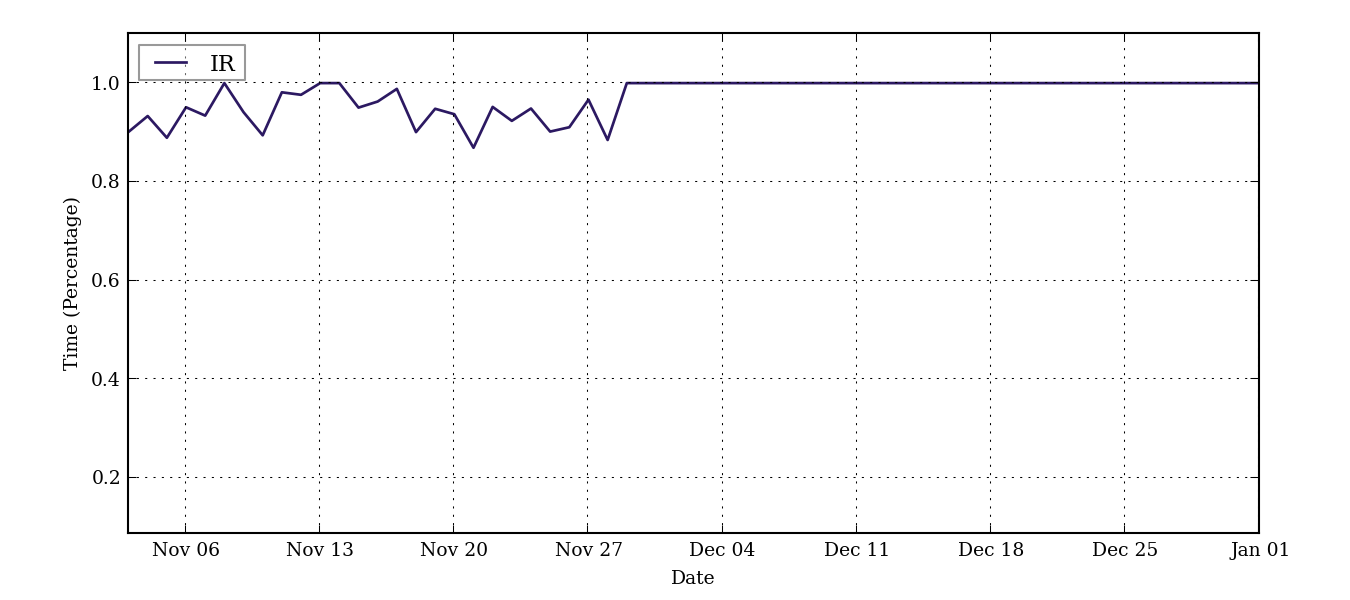}\hfill
    }
    \caption{Suspected Throttling Event, Nov 2011 - Dec 2011}
    \label{findings:throttling_events:nov_2011}
\end{figure}

Figures \ref{findings:throughput} and \ref{findings:variance} outline the periods of time where fluctuations in values and variance of throughput exceeded thresholds, respectively. The performance measurements and indicators outlined in the prior section during these two extended periods of interest are documented in the graphs of Figures \ref{findings:throttling_events:oct_2012} and \ref{findings:throttling_events:nov_2011}.

\begin{figure}
\centering
\subfloat[Throughput (Daily)]{
    \renewcommand{\arraystretch}{1.4}%
    \begin{tabular}{| l  c  l | c |}
    \hline
        Start Date &  & End Date & Change \\ \hline
        \multicolumn{4}{|c|}{Likely Events} \\ \hline
        2011 Nov 30 & - & 2012 Aug 15 & -77.0\% \\ 
        2012 Oct 4 & - &  2012 Nov 22 & -69.0\% \\ 
        \hline \multicolumn{4}{|c|}{Possible Events} \\ \hline
        2010 Jan 29 & - & 2010 Feb 2 & -56.0\% \\
        \multicolumn{3}{|c|}{2010 Mar 21}  & -36\% \\ 
        \multicolumn{3}{|c|}{2010 May 12}  & -32\% \\ 
        \multicolumn{3}{|c|}{2010 Sept 16} & -39\% \\
    \hline
    \end{tabular}
    \label{findings:throughput}    
}
\subfloat[Variance (Weekly]{
    \renewcommand{\arraystretch}{1.4}%
    \begin{tabular}{| l  c  l | c |}
        \hline
        Start Date &  & End Date & Change \\ \hline
        \multicolumn{4}{|c|}{Likely Events} \\ \hline
        2011 Nov 28 & - & 2012 Aug 20 & -98.0\% \\
        2012 Oct 01 & - & 2012 Dec 3 & -82.0\% \\
        \hline \multicolumn{4}{|c|}{Possible Events} \\ \hline
        2010 Feb 15 & - & 2010 Feb 22 & -92.0\% \\
        2010 Mar 19 & - & 2010 Mar 26 & -48.0\% \\
        2010 May 10 & - & 2010 May 24 & -68.0\% \\
        2010 Oct 25 & - & 2010 Nov 01 & -68.0\% \\
        2011 Jan 31 & - & 2011 Feb 7 & -75.0\% \\
        2011 Jul 25 & - & 2011 Sept 5 & -78.0\% \\
        2012 Dec 24 & - & 2013 Jan 21 & -70.0\% \\
    \hline
    \end{tabular}
    \label{findings:variance}
}
\caption{Precipituous Changes of Service Quality in Iran, 2010-2012}
\label{findings:tablesvalues}
\end{figure}

Although our methodology has not taken a deterministic view from prior awareness, these two major events mirror our prior understanding of periods of disruption. While we could not find instances where M-Lab or similar tests were used in the commission of news reports on Iran's Internet, our results often mirror claims such as ``The Internet In Iran Is Crawling, Conveniently, Right Before Planned Protests''\cite{tnw:crawling}. We also find potential events as detected by changes in performance surrounding holidays, notable protests events, international political upheaval and important anniversaries, such as Nowruz, the Arab Spring and 25 Bahman (Persian calendar date, early-to-middle February). These also often parallel more overt forms of disruption, such as the filtering of secure Google services (September 24 - October 1 2012) and significant jamming of international broadcasts (January 31 - February 7 2011, October 2012).

Figure \ref{findings:tablesevents} demonstrates the reverse methodology, correlating reported incidents of public protests or opposition rallies with NDT measurements.\footnote{We identify as incidents of public protests, the following dates: 2010-02-11, 2010-03-16, 2010-06-12, 2011-02-14, 2011-02-20, 2011-03-01, 2011-03-08, 2012-02-14, 2012-10-01} The dates identified focus on country-wide mobilizations, rather than localized events such as protests by Iran's Ahvaz Arab minority population in Khuzestan during April 2011. While it is probable that localized throttling occurs, due to limitations in sampling, it may not be possible to detect such actions until more nodes of measurement are available. In addition to the throughput for the primary or initial day of the event, the table identifies the data trends of the week and the time period of the month before and after. This timeframe takes into consideration the measures applied by the state to stifle mobilizations for publicly-announced events and mitigate further political activities. These reactions are reflected both online and offline, during  Winter 2011, in reaction to plans to protest on 25 Bahman, former Presidential candidates Mir-Hossein Mousavi and Mehdi Karroubi were detained and kept in house arrest by security forces for their role in the reformist politics. These mobilizations are compared against a two month window as a baseline of the general capacity of the network at that time.

We find a direct correlation between the precipituous decline in connectivity for February 2012's anniversary of the detention of Mousavi and Karroubi, as well as October 2012's currency protests. During all events, variation between the average RTT of two month mean and those recorded during the event stayed within a 20\% threshold. While we anticipated 10 February 2010 as a potential throttling event, we find that three days later, the date of our minimum measurement during the period, was associated with both street protests and the hacking of opposition news sites \cite{rsf:internetenemies}. Similarly, while our weekly minimum for the first anniversary of the 2009 Presidental election falls on the anticipated day, the decrease in performance does not exceed a reasonable threshold of variation. Finally, March 2010 stands out as a strong case of a negative correlation across the two-month context, however, it may be reasonsable to consider the patterns of network use established previously and this period's proximity to the Nowruz holiday. Contrarily, if lower-end consumer users are subjected to more aggressive throttling, and as a result decide stay off the Internet until speeds improve, the median of the national throughput would increase. Against our two-month baseline, there was a 16\% decrease in NDT clients on May 16 and a weekly minimum of 41\% decrease. Thus, this undergirds the need to detect both precipitous decreases \textit{and increases} in performance.

\newcolumntype{L}{>{\bfseries}l<{}}
\begin{figure}
\renewcommand{\arraystretch}{1.3}
\centering
\begin{tabular}{|L|c|L l l|c|c|c|}
\hline    
Event &  Day Of &  & Wk-Min & &  Wk-Mean & 2-Month \\ \hline
2010-02-11 & 0.18 & 0.14  & 2010-02-14 & -34.3\% & 0.20 & 0.19 \\ \hline
2010-03-16 & 0.26 & 0.19  & 2010-03-13 & +7.3\% & 0.22 & 0.17 \\ \hline
2010-06-12 & 0.16 & 0.16  & 2010-06-12 & -21.6\% & 0.19 & 0.20 \\ \hline
2011-02-14 & 0.18 & 0.15  & 2011-02-17 & -18.9\% & 0.18 & 0.18 \\ \hline
2011-02-20 & 0.22 & 0.15  & 2011-02-17 & -21.1\% & 0.20 & 0.18 \\ \hline
2011-03-01 & 0.18 & 0.12  & 2011-03-04 & -52.3\% & 0.17 & 0.19 \\ \hline
2011-03-08 & 0.16 & 0.16  & 2011-03-08 & -14.0\% & 0.18 & 0.19 \\ \hline
2012-02-14 & 0.03 & 0.03  & 2012-02-14 & -102.9\% & 0.07 & 0.06 \\ \hline
2012-10-03 & 0.25 & 0.09  & 2012-10-04 & -86.2\% & 0.20 & 0.16 \\ \hline
\end{tabular}
\caption{Incidents of Widespread Protests and Median Throughput in Iran, 2010-2013}
\label{findings:tablesevents}
\end{figure}

We anticipate based on the network principles and diurnal patterns previously established that network load will be directly correlated with higher roundtrip times. In Figures \ref{findings:throttling_events:oct_2012:rtt} and \ref{findings:throttling_events:nov_2011:rtt}, there appears to be no such relationship between the round trip time of the clients' traffic and measurements of service quality. It is less likely that these changes were the product of heavy use.

Generally, sudden drops in service quality can be attributable to changes in domestic networks or the availability of upstream providers, due to a multitude of factors such as physical damage or electronic attacks. During the interval of throttling identified as the beginning of October 2012 and described in Figure \ref{findings:throttling_events:oct_2012}, the main international gateway provided by Information Technology Company (AS12880), experienced routing failures to networks connected through Telecom Italia Sparkle (AS6762)\cite{renesys:bulletin_oct_2012,renesys:bulletin_oct_2012_2}. However, Iran's international gateways are amongst the most unstable on the Internet, with frequent, short periods of routing failures even during normal operations \cite{nanog:bgp_update}. Therefore, it is important to differentiate relatively routine failures from protracted and wide-cutting outages. Additionally, as latency is a partial product of the psychical distance of a network path, changes in distances of traffic traversing paths to the global Internet should show as changes in latency. It remains unclear whether these reported disruptions were due to connectivity failures, or downtime due to maintance and application of changes to the network. In the case of the October currency crisis event, these reported failures were short-lived as normal service appears to have been restored within minutes, and little change in latency is measured. 

Despite the centralization of domestic peering through the key points of control, Iran has a diversity of physical pathways connecting the country to the global Internet, creating upstream redundancy. A clear demonstration of the effect, and minimal impact for our purposes, of infrastructural failure occurs within our October 2012 event, when an attack against a natural gas pipeline by the Kurdistan Workers Party caused damage to infrastructure providing connectivity through Turkcell Superonline \cite{renesys:blasts}. While these changes were detected in Figure \ref{findings:throttling_events:oct_2012:throughput}, they were within the tolerance levels established during the ongoing throttling event. This event also appears to be reflected in a marginal increase of the average and minimum round trip times of Figure \ref{findings:throttling_events:oct_2012:rtt}, as clients compete over diminished resources and traverse potentially longer routes to M-Lab servers.

Thus far, we have primarily focused on two extended periods of time for analysis in order to explore the technical metrics outlined in Section \ref{sec:setup} in an environment subject to false positives. In Figure \ref{findings:tablesvalues}, we enumerated those events that triggered our detection mechanism, including shorter term periods without vetting their veracity. As discussed throughout the paper, the deeper we narrow our evaluation to a network-level granularity, the more subject we are to the limitation of the manner in which our data was collected. The longer a detected abnormality lasts, the higher confidence we can assert our results are not aberrant testing, that independent mechanisms are causing peculiarities in network that should not otherwise occur.

Applying these lessons, it would appear that a number of false positives, general the result of wide fluctuations in measures, trigger the detection mechanism of one metric but do not register on elsewhere. We remain especially interested in the reported incidents February 2010, March 2010, Feburary 2012, and January 2013. Additionally, we manually identify early April 2010 as an interesting period based on low variance between ISPs, the amount of time spent in a network-congested state and a rapid change in the daily variance of throughput measurements on the top five networks. Other remaining periods of interest consist of short timeframes that bear noteworthy links between metrics, but we cannot confidently assert are meaningful, including several periods in June 2010, late October 2010 and July - August 2011. 

\begin{figure}
 \centering
  \includegraphics[width=\textwidth]{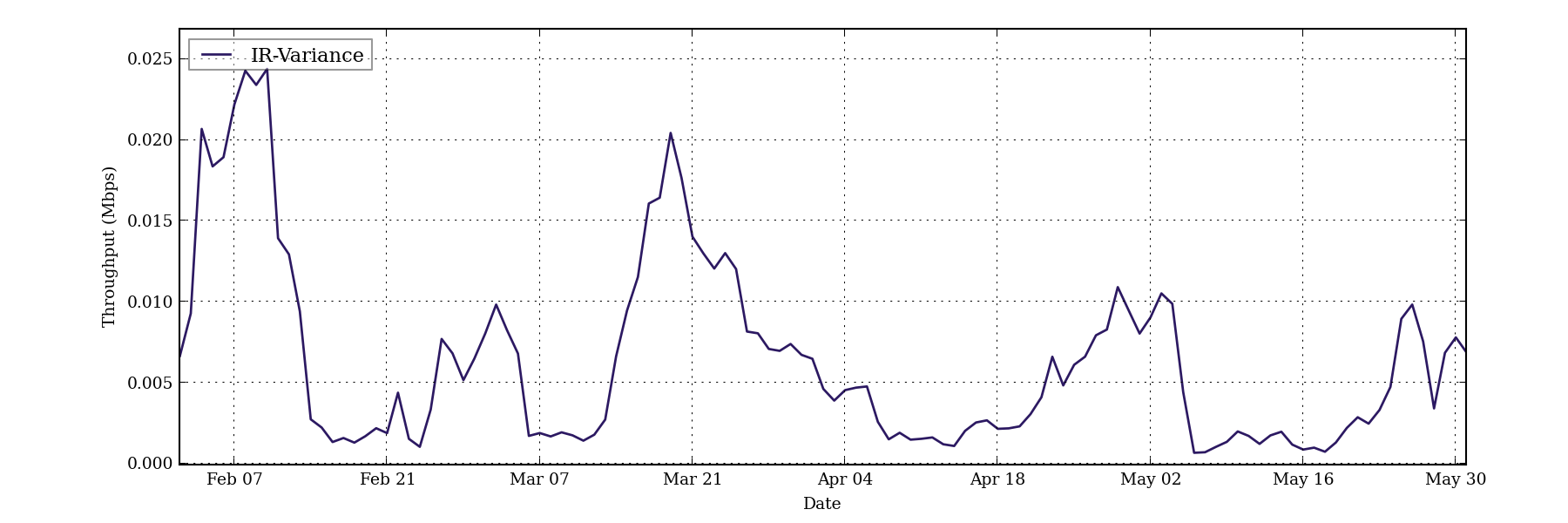}
 \caption{Throughput Variance Amongst Iranian ASNs, February 2011 - May 2011}
 \label{fig:nov11:var}
 \end{figure}

\subsection{Control Groups}\label{sec:controlgroups}

While throttling and disruption of international connectivity may be useful for intermediaries that intend to stifle the free flow information for the general public, Internet-based communications have grown to be a core component of state operations, diplomatic functionalities and international business transactions. The identification and segmentation of critical networks provides a means to mitigate the deleterious effects of communications loss. Traffic prioritization or exemptions to disruptions for white-listed users and protocols should be trivally easy in any modern network appliance. The utilization of such features for non-censoring purposes on Iranian networks is already well documented \cite{hassani2010qos}. Therefore, we expect that high value networks, such as government ministries and banks, would potentially be spared the majority of disruptions if possible. 

Conversely, from the perspective of inferring meaning from raw data, we can posit a diversity of circumstances for why a particular network or client would be less aversely affected by country-level disruptions. Since all connections appear to route through the same intermediaries, those that continue to have normal service likely have been purposefully excluded. However, it remains unclear as to whether throttling occurs at the international gateway or is left to be implemented by the service providers. It may hold that throttling is mandated administratively through the legal authority of the telecommunications regulators, but implemented technically by the service providers. Comments from former staff of Iranian ISPs have indicated that bandwidth restrictions have previously been enforced through an order delivered over the phone or by fax. These claims have gone on to allege that some ISPs, generally smaller and regional providers, delay or limit compliance as they attempt to balance the demands of the state with the possibility of losing customers over poor quality of service.\footnote{We note these anecdotal stories as a theoretical condition capable of being tested, not as evidence or an asserted mechanism of implementation. Publicly-disclosable documentation of the ISP role in censorship from either tool-makers or former staff has been difficult to source due to the security issues.} If certain ISPs are more likely to delay implementation of throttling orders, this may add an indicator to our detection and establish performance disruptions based on intent. In such a scenario, after a mandate is distributed, we would anticipate seeing that larger ISPs enter into periods of throttling before smaller providers.

\begin{figure}
 \centering
    \includegraphics[width=\textwidth]{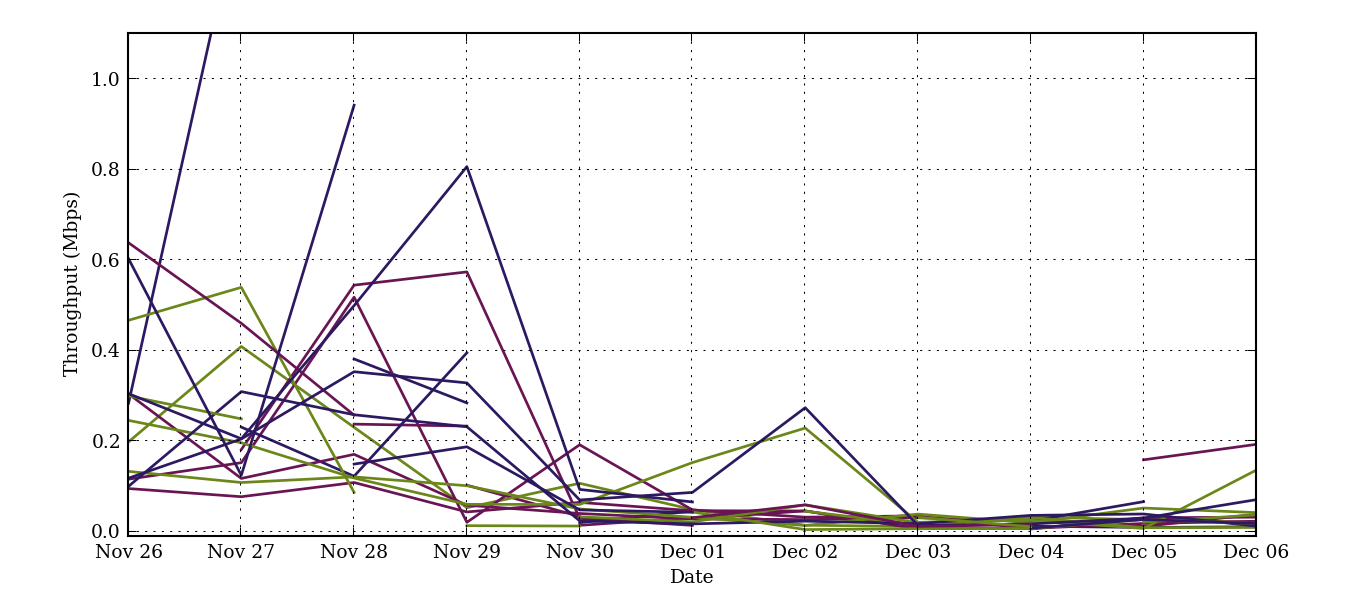}
    \caption{Throughput, Aggregated based on ASN, November 2011 - December 2011}
    \label{findings:throttling_events:nov_2011:asn}
 \end{figure}

Were the granularity of our dataset to allow for it, this question would potentially be answered by demonstrating that the majority of networks witnessed changes within a very close promixity of each other due to central coordination of implementation, as opposed to the delays and differing interpretations that may come with diffuse, independent implementation. This would likely require a data source that is statistically valid when reduced to an hourly basis, rather than our daily aggregation in Figure \ref{findings:throttling_events:nov_2011:asn}. Additionally, this hypothesis requires prefix-level evaluation, and thus inevitably runs into the limitations of passive, crowdsourced datacollection, as smaller networks will have fewer users running NDT tests, thus rendering assessment less truthworthy or responsive. This claim also assumes that smaller ISPs are not subject to the throttling of upstream domestic peers.

\begin{figure}
 \centering
 \subfloat[Normal Period, February 2011 - November 2011]{\includegraphics[width=\textwidth]{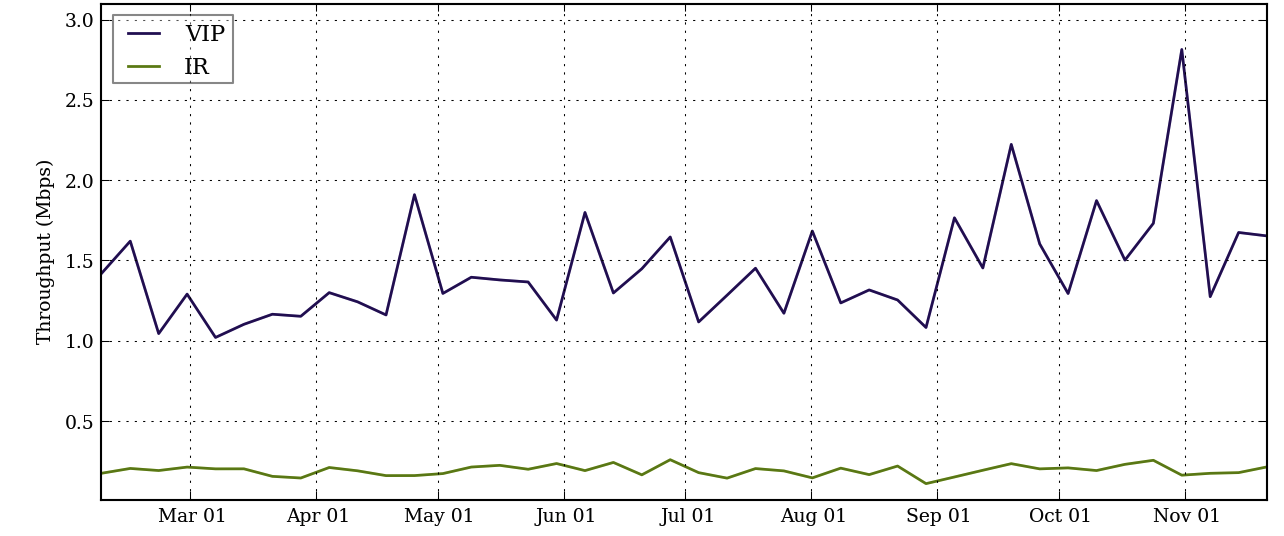}\label{fig:vip:comparative:1}}\hfill
  \subfloat[Throttling Event, November 2011 - June 2012]{\includegraphics[width=\textwidth]{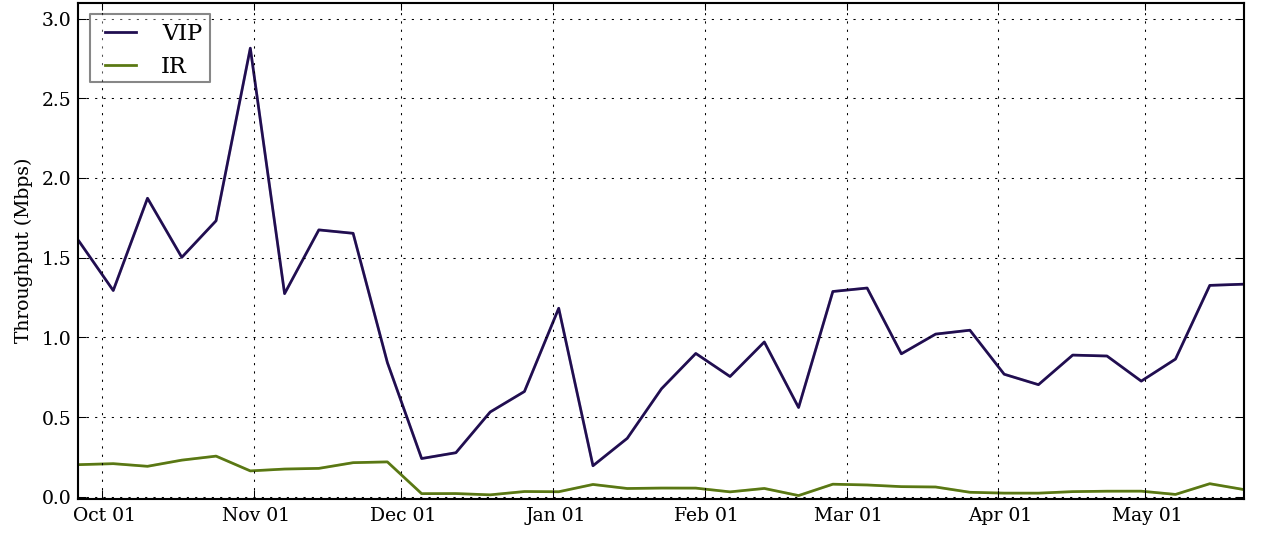}\label{fig:vip:comparative:2}}
 \caption{Comparative Throughput}
 \label{fig:vip:comparative}
 \end{figure}

As M-Lab's NDT data contains the IP address for clients, we are able to identify those networks within higher tiers of performance, which could then be used to create a control group or baseline in order to track service changes. Using the two significant throttling events described in Section \ref{sec:throttling_events}, we then identify networks based on a threshold of 95\textsuperscript{th} percentile of throughput rates, and thus networks that may have priority during disruptions. In order to test this hypothesis, we compare the median throughput rates of our privileged subset against the national median. Since the sample of clients is based on a minority of domestic Internet users and more susceptable to fluctations or misattributions, we are more interested in the trends of these users and the identification of the types of networks they belong to. Figure \ref{fig:vip:comparative} demonstrates the relationship between these higher tier services before and after a suspected throttling event. Based on our assumptions of the narrowness of exemption rules, these clients are aggregated within the most restrictive IP prefix available through Team Cymru's IP to ASN Mapping service \cite{cymruip}.

Appendix Figures \ref{appendix:vip:before} and \ref{appendix:vip:during} enumerate the number of clients that recorded measurements within the higher percentiles for the country, based on address prefix. As the number of addresses in a prefix varies according to how they were assigned originally, the immediate value of such data is limited. A proper evaluation of the trends of networks requires an understanding of scale and utilization of the network. For example, Parsonline's \textit{91.98.0.0/15} and \textit{91.99.32.0/19} address prefixes are large consumer IP pools of over 135,000 addresses, across a range of connectivity methods and customer types. It would therefore be less interesting if these networks contained a substantial number of high percentile clients than if a smaller ISP with less customers or peculiar ownership to perform well. Also within this set is Mobin Net, the nationally licensed WiMAX data monopoly, which appears to provide service from 128kbps to 2Mbps packages, far beyond most ADSL offerings.

\begin{figure}
\centering
\begin{tabular}{|l|l|c|c|c|c|}
\hline
ASN & Owner & $\Delta$ & $\Delta$ (+2) & $\Delta$ (+10) & Oct 2012 \\
\hline
AS12660 & Sharif University of Technology& -74.64\% & -70.46\% & -2.43\% & -58.62\% \\
AS12880 & Information Technology Company (ITC) & -95.77\% & -93.26\% & -84.94\% & -91.57\% \\
AS16322 & Parsonline & -94.26\% & -91.83\% & -67.05\% & -86.46\% \\
AS25124 & DATAK Internet Engineering & -90.74\% & -93.42\% & -76.66\% & -87.23\% \\
AS25184 & Afranet & -87.73\% & -78.46\% & -32.25\% & -68.23\% \\
AS29068 & University of Tehran Informatics Center & -79.99\% & -90.31\% & -47.37\% & -69.43\% \\
AS31549 & Aria Rasana Tadbir & -94.46\% & -93.19\% & -82.86\% & -91.60\% \\
AS39074 & Sepanta Communication Development & -89.39\% & -90.92\% & -75.06\% & -91.60\% \\
AS39308 & Andishe Sabz Khazar & -90.34\% & -76.92\% & -82.14\% & -80.96\% \\
AS39501 &  Neda Gostar Saba Data Transfer Company & -94.29\% & -89.80\% & -70.38\% & -86.13\% \\
AS41881 & Fanava Group & -79.30\% & -83.64\% & -83.98\% & -73.19\% \\
AS43754 & AsiaTech Inc. & -89.12\% & -89.49\% & -82.57\% & -86.36\% \\
AS44244 & Irancell & -87.68\% & -88.40\% & -69.52\% & -77.57\% \\
AS44285 & Shahrad Net Company Ltd. & -91.81\% & -85.17\% & -80.06\% & -62.23\% \\
AS48159 & Telecommunication Infrastructure Company & -94.72\% & -94.76\% & -89.54\% & -87.06\% \\
AS49103 & Asre Enteghal Dadeha & -95.51\% & -91.48\% & -71.45\% & -69.02\% \\
AS50810 & Mobin Net Communication Company & -95.50\% & -94.63\% & -80.26\% & -91.10\% \\
\hline
\end{tabular}
\caption{Recovery After Throttling Event (Nov 2011)}
\label{finding:nov2011_asn_diff}
\end{figure}

Thus, for our purposes in assessing the trends of networks, we rely on statistical measurements of relative changes. In the interest of stronger sampling, we rely on larger ASN aggregation and define a threshold for consideration to those that have performed measurements for at least half as many days as the time period. Figure \ref{finding:nov2011_asn_diff} demonstrates the recovery of network throughput by comparing the mean values of the two month period preceding the November 2011 incident with, 
\begin{inparaenum}[i)]
    \item the two months immediately after,
    \item February to April 2012,
    \item August to October 2012,
    \item the comparative degradation of performance during the October 2012 event.
Accordingly, between the two months preceding and the first two month following the November 2011 event, every network under consideration experienced more than a 74\% drop in througput. Even within those networks (ASNs) that do not meet our qualifications, only one experienced an increase in throughput performance immediately after the November 30 2011 disruption, the prefix 80.191.96.0/19 run by the ITC, which according to reverse DNS records on the block appears to provide commercial hosting services and connectivity for academic institutions, such as Shiraz University. 

\begin{figure}
 \centering
  \includegraphics[width=\textwidth]{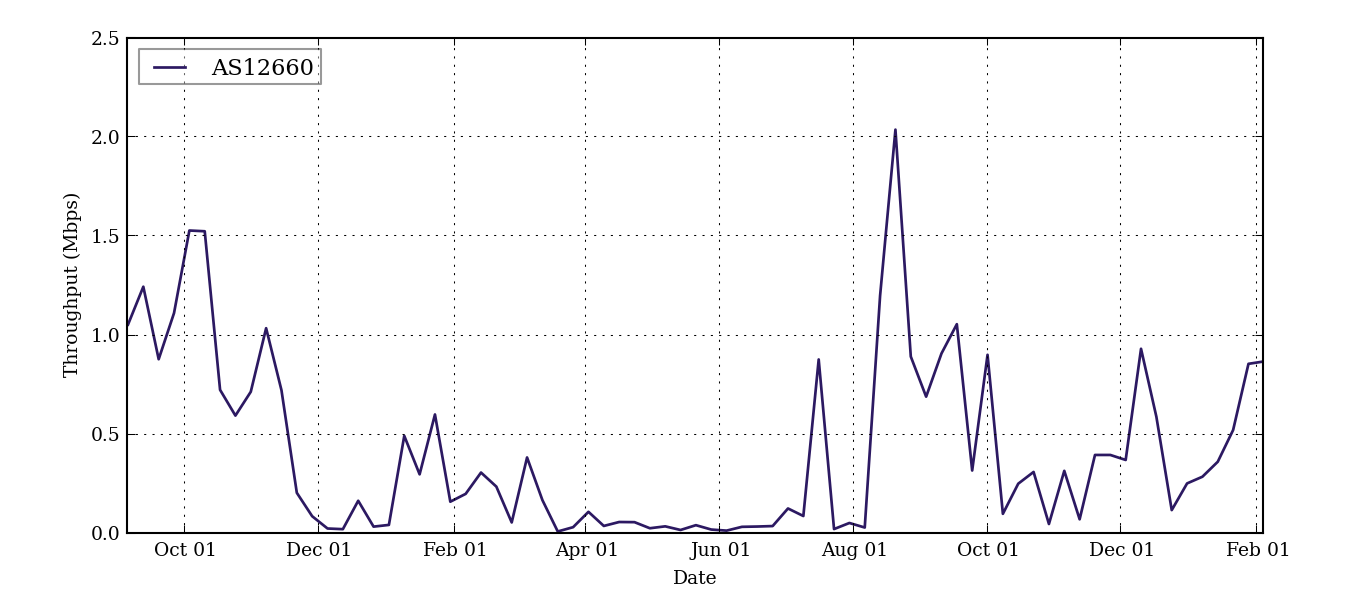}
 \caption{Throughput for Sharif University (AS12660), Oct 2011 - Jan 2013}
 \label{fig:sharif_throttle}
 \end{figure}

  \end{inparaenum}

While no major network appeared to have escaped the events of November 2011 or October 2012, clear trends exist in the level of disruption and the rate of return to normality. Academic institutions, as evinced in Figure \ref{appendix:vip:before}, have historically had access to faster connectivity prior to the November event \cite{upen:faa}. Figure \ref{fig:sharif_throttle}, the trends for Sharif University of Technology, demonstrates that while the university has been significantly affected by network degradation, it recovered faster than other networks. The networks owned by the Fanava Group, University of Tehran Informatics Center, and Sharif University were the only three to experience less than a 80\% decline in throughput. Eight months later, only Sharif, University of Tehran and Afranet had begun to return to their normal levels. This mirrors Figure \ref{fig:vip:comparative}, where the trend of the mean of the $95^{th}$ percentile of tests was impacted by the throttling at the initial event. However, this subset recovered to an approximation of prior values more quickly than others.

Considering its large consumer dialup and ADSL Internet offerings, on first glance Afranet's strong recovery is unexpected in comparison to other ISPs, such as Parsonline. While there are no upfront indicators of the type of connection used for a specific client, it is possible to infer the nature of the source or the network it is associated with, from indirect means. First, the registration and announcement of routing information may provide labeling of the use, such as ``Shabake Almas Abi,'' hinting that the client originated from a smaller ISP that Afranet is the upstream provider for, or ``AFR@NET company, Tehran, Dialup pool,'' which is most likely a consumer address pool. When we perform reverse DNS queries on the address and prefix, the answers may reveal the owner of the network through the domains pointing to the space. However, any prefix returning more than a marginal number of responses likely indicates that the network is used for commercial purposes, as consumers generally do not host sites on home connections, especially where addresses change dynamically. In our recover period of August to October 2012, the majority of clients performing NDT tests on Afranet originated from the blocks of 217.11.16.0/20 and 80.75.0.0/20, which appear to host the infrastructure or provide connectivity for prominent commercial entities, such as Iranohind and Saipa Automotive, or 79.175.144.0/20 and 31.47.32.0/20, which we suspect to be smaller ISPs or hosting providers. No clients originated from the home and dial up address pools that were identified at 78.109.192.0/20 and 79.175.176.0/24.

\section{Conclusions and Further Questions}\label{sec:conclusions}

Absent independent, quantative evidence of claims, Iranian public officials have argued that, ``in spite of negative ads and fallacies \ldots recent numbers prove that Internet speed is very satisfactory in Iran,'' defying the everyday experience of the public \cite{citlab2012}. In this paper, we sought to establish a historical, quantitative dataset used to describe a phenomenon that thus far has existed solely in the realm of rumors and anecdote. Immediately upon the most shallow evaluation of the trends, we find frequent and prolonged changes to the service quality of clients originating from Iran. We attempt to account for these changes based on more quotidian explanations of upstream connectivity degradation, domestic infrastructure failure, or increased network traffic. While we do find noteworthy incidences of publicly-reported network outages and diurnal patterns of service quality, these do not account for the length and timing of disruption, or the extent of impact. 

In order to test the hypothesis under consideration, we present a number of testable assumptions about how artificial throttling would manifest within our measurements, grounded in an understanding of the technological and administrative principles at work. While we quickly run into frustrations arising from the scope and breadth of our dataset, we are also able to derive an initial set of answers. When we are limited in the confidence of results due to the sample size, origin or consistency of information, then we can narrow our investigation based on the correlation of multiple analyses. Rather than detecting based on simple indicators of throughput or variance, we are required to look at a range of measurements.

By its nature, throttling is opaque occurrence and technical measurements can rarely infer intent, however, the service disruptions documented herein cannot be accounted for within normal expectations of network operations. We establish that the periods of disruption identified are widely applied across all networks but vary in magnitude and recovery, lasting from only a few days to several months. These differences parallel the purpose of the networks, thus implying special consideration of the socioeconomic impact in application of the disruption.


Finally, apart from citing specific historical and infrastructural circumstances, this paper attempts to describe the first steps of a broader framework to account for the stability and accessibility of international connectivity in states that impose limitations on free expression and access to information. As noted by others, Iran is within a large cohort of countries that have centralized international communications transit to a limited set of gateways.

\subsection{Remaining Questions}

In order to continue the development of such a monitoring and accountability tool, we anticipate the integration of NDT tests with independent sources of complementary data and solicit input toward a number of outstanding questions:

\begin{itemize}

\item \textbf{What remaining TCP/IP and NDT indicators apply to throttling?}

Thus far, our indicators have relied largely on a few metrics available from NDT that we anticipated would be directly associated with disruption. This subset does not represent the full suite of measurements and properties available from M-Lab. Of particular interest remains the raw network stream captures retained from the individual NDT tests. We anticipate using the \textit{time to live} (TTL) IP property, a counter that decrements for each router it traverses in order to detect routing loops, to detect changes in the network, as well as monitoring fields that may be manipulated by an intermediary attempting to prioritize traffic, such as the ToS field.

\item \textbf{What could we learn from vendor documentation?}

The principles undergirding our hypothesis and analysis have largely been derived from the documentation available for the Linux \textit{traffic control} subsystem, which provides the means for the Linux operation system to perform throttling and shaping on egress network traffic. This documentation is especially useful given the wealth of peer-reviewed publications on its implementation concepts, configuration and performance, as well as the proliferation of Linux-based devices at the core of modern public networks. Similar features exist for Cisco devices, under quality of service traffic classes such as \textit{rate-limit} and \textit{traffic-shape} \cite{ciscoqos}, and for Huawei within the \textit{traffic behavior} definitions \cite{huaweiqos}. Few telecommunications equipment vendors produce network stacks that are written from scratch or ignore the basic principles used for managing traffic flows, such as `token bucket' mechanisms. Furthermore, open source reporting indicates that equipment from both of the prior mentioned manufacturers are core to Iran's Internet \cite{wapo:intranet}. Therefore, performance testing, emulation of environments, identification of instruments of implementation and the development of more precise methodologies that are closer to real world conditions can potentially be accomplished using off-the-shelf equipment or documentation if the quality of data allows. 

\item \textbf{What is the correlation between the disruption events and the measured increase of packet loss or latency?}

We have previously asserted that the decrease in performance without a corresponding increase in loss or latency constitutes an abnormal network condition that may represent the imposition of rate limits. However, we do not hold directly that throttling cannot be associated or accomplished through these means. We hold both scenarios as loose relationships, and infact the literature on \textit{traffic control} describes artifical loss as a means of throttling. Routers may police the rate of traffic flows by dropping packets once an assigned buffer is filled. Thus, the relationship between throttling and the factors of loss and latency is nebulous. In the case under consideration, the former is particularly pressing. While during three days in January and February 2010, aggregated measurements register a 100\% increase in round trip time within twenty four hour periods, this instability is unmatched within our three year dataset and potentially dismissible due to the small sample size during that time. The rapid degradation of Iran's connectivity in late November 2011 is associated with over 30\% packet loss, a nearly 1000\% increase over the preceeding days, only a few days within this period registered less that 10\% loss. The October 2012 throttling event fits this theme with a consistent rate of about 10\% loss. The only metric of latency that matches the throttling event is the pre-congestion round trip time, which registers a 49\% increase, and maximum RTT, -10.5\% decrease, in November 2011, but is not paralleled by the measurements of average and minimum time that we focus on.

\item \textbf{At which level of network infrastructure does throttling occur?}

As we note throughout discussions on Iran's infrastructure and our findings, rumors and references in public documents have pinpointed some level of throttling occuring on the part of the end-consumer Internet providers. This is to be expected considering that ISPs retain legal responsibilities to police criminal content, with the TCI serving an auxiliary role running its own filtering and deep packet inspection. Thus far, NDT has not clearly provided the granularity of data required to answer this question under the rubric outlined in Section 5.2. Although these records show variations in the extent of disruption, at any level in the path, an administrator would be able to differentiate the rules for handling traffic from different networks. These questions also reflect the frustrations in determining the most narrow application of exemptions to disrupt, as we search for IP prefixes, ASN and even cities that have been less aversely affected by disruption. 

\item \textbf{Is the technical application of disruption rules consistent across domestic ISPs, instances of throttling, and countries?}

Thus far we have avoid asserting a set of properties that we believe are direct and exclusive evidence of intermediary throttling. We have noted at length the difficulties of establishing such confidence within the nature of the test and the opacity of administration. However, we also anticipate variances in implementation. Between any service provider or, more expansively, countries, differences of technical capacity and infrastructure will lead to different opportunities or approaches. The example of Bittorrent throttling by American and European ISPs provides an independent and more thoroughly investigated illustration of the  diversity of means available to manipulate the connections that pass through a network, and demonstrates the role of specific equipment in various strategies. Additionally, intent matters. If an intermediary is primarily concerned with disrupting streaming media, Internet telephony or anti-filtering connections, then dropping packets or sending connection resets may be a more efficient approached. Moreover, in such a case, ISPs may even offer their users an initial burst of fast speeds, that are then throttled down. Thus, a universal and explicit formula, as opposed to statistical inference and manual inspection, is unlikely to ever be possible.

\item \textbf{What is the most appropriate criteria to filter tests for consistency?}

We impose conditions regarding the length and integrity of the records used in our assessment, in order to filter out misleading or error prone measurements. These limitations are: tests that lasted longer than 9 seconds and less than an hour, and exchanged at least 1 packet and less than 120,000 packets. Additionally, it may be useful to impose restrictions based on the M-Lab server, for purposes of consistency in routes and network conditions. However, such a decision would require tests to ensure this does not impede m-labs ability to accommodate changes in international routes. We do not consider upstream tests solely for the sake of brevity, although this direction may be equally important should the throttling of upload connect be a means of curtailing the outward flow of media.

\item \textbf{What is the future relevance of NDT in monitoring the next generation of throttling?}

Iran's strategies of censorship have followed a historical trend of increasing precision of disruption. Whereas the June 2009 elections corresponded with a multiple week outage of SMS services, by early Spring 2013 keyword filtering on political slogans or terms associated with controversial issues had become a normal occurrance. The blocking of SSL in February 2012 had shifted to the blocking of SSL to selection networks and the redirection of secure traffic through the interception of DNS requests [IIIP 1] . Similiarly, reports of throttling appear more specific, such as SSL or multimedia traffic in general, or SSL to services such as Google [IIIP 3]. The more that a censoring intermediary narrows its understanding and limiting of offensive traffic to a `black list,' a strategy that has substantial political and economic value, the less that these disruptions will be reflected within the NDT test. However, a countervaling pressure is presented by the adoption of sophisticated strategies by anti-filtering tools to disguise or randomize their network traffic in a manner that makes deep packet inspection increasingly difficult and costly, such as the obfsproxy mechanism employed by Tor and Psiphon. Without the ability to confidently distinguish normal traffic from privacy-perserving connections that it cannot control, censors may be forced back into more to the broad throttling regime or shifting to a `white list' strategy that NDT would detect. 

\item \textbf{How do we best filter tests for consistency and accuracy?} We limit the records used in our assessment in order to filter out misleading or error prone measurements. Limit the measurements to those downstream tests that lasted longer than 9 seconds and less than an hour, and exchanged at least 1 packet and less than 120,000 packets. Additionally, it may be useful to further limit based on the m-lab server, for purposes of consistency in routes and network conditions. However, such a descision would require tests to ensure this does not impede m-labs ability to accomodate changes in international routes. We do not consider upstream tests solely for the sake of brevity. This direction may be equally important should the throttling of upload connect be a means of curtailing the outflow flow of media.


\end{itemize}

\section{Acknowledgements}

This research would not have been possible if not for the substantial contributions of a number of individuals who I am privileged to even know and deeply regret not being able to acknowledge in name; the chilling effect of censorship and state intimidation is not limited to the borders of a country. Fortunately, it is possible to recognize Briar Smith, Meredith Whittaker and Philipp Winter for their technical, professional and moral support, making this matter of curiosity into something more professional, which I hope contributes to the ability of the public to protect such an important medium of expression.

\bibliographystyle{plain}
\bibliography{paper}
\newpage
\section{Appendix}\label{sec:appendix}
\begin{figure}
    \centering
    \subfloat[November 24 2011]{
        \includegraphics[width=.5\textwidth]{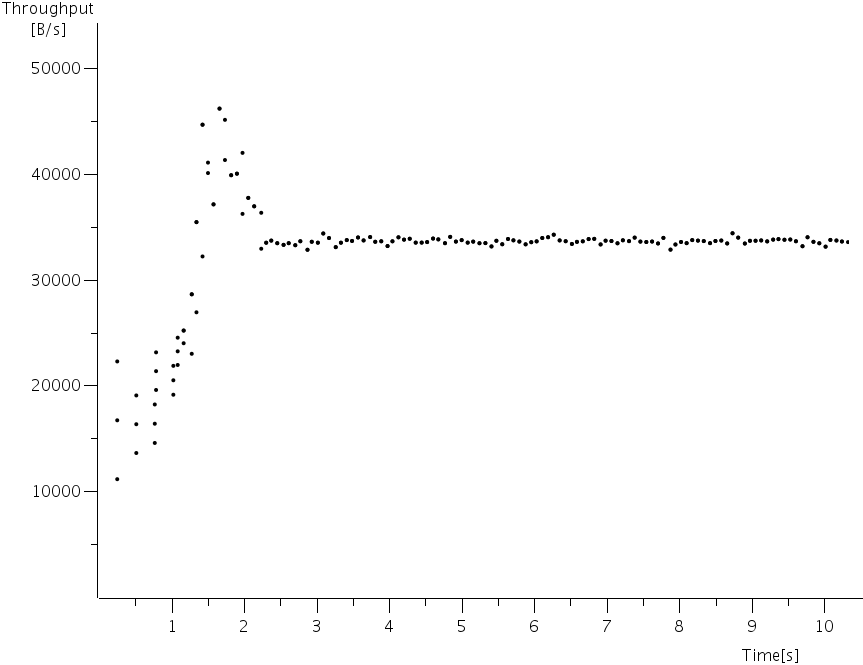}\hfill
    }
    \subfloat[December 4 2011]{
        \includegraphics[width=.5\textwidth]{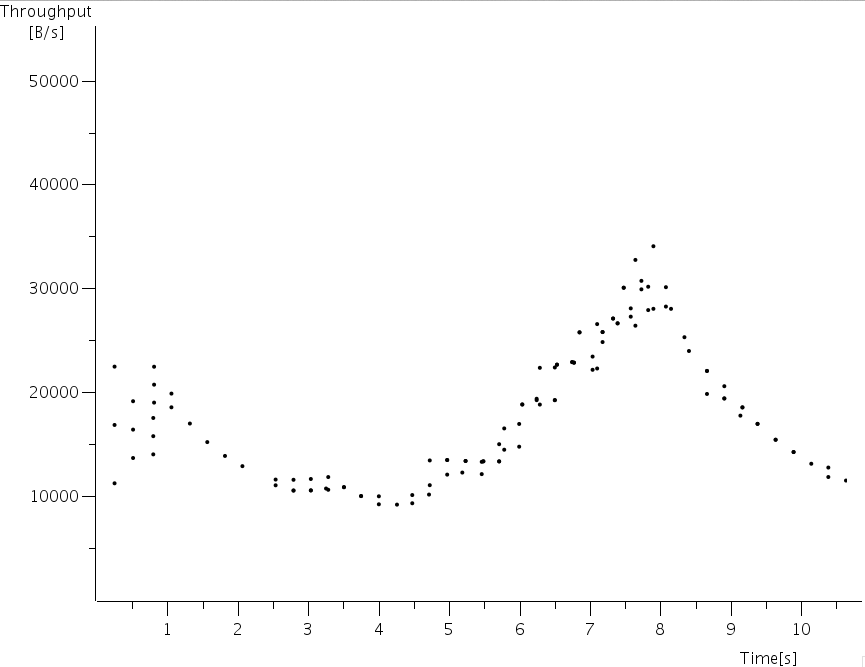}\hfill
    }
    \caption{TCP Stream Throughput, Before and During Throttling}
    \label{findings:throttling_events:tcpdump}
\end{figure}
\begin{figure}
\begin{tabular}{| l | l | c |}
\hline
SHARIF-EDU-NET Sharif University of Technology, Tehran,Iran & 213.233.160.0/19 & 165 \\ \hline
NGSAS Neda Gostar Saba Data Transfer Company & 188.158.0.0/16 & 126 \\ \hline
UT-AS University of Tehran Informatics Center & 80.66.176.0/20 & 112 \\ \hline
SHARIF-EDU-NET Sharif University of Technology, Tehran,Iran & 81.31.160.0/19 & 73 \\ \hline
ASIATECH-AS AsiaTech Inc. & 79.127.32.0/20 & 60 \\ \hline
PARSONLINE PARSONLINE Autonomous System & 91.98.0.0/15 & 53 \\ \hline
IR-ASRETELECOM-AS Asre Enteghal Dadeha & 188.34.0.0/17 & 49 \\ \hline
NGSAS Neda Gostar Saba Data Transfer Company & 188.159.0.0/16 & 48 \\ \hline
PARSONLINE PARSONLINE Autonomous System & 91.99.0.0/19 & 45 \\ \hline
PARSONLINE PARSONLINE Autonomous System & 91.99.0.0/16 & 42 \\ \hline
DNET-AS Damoon Rayaneh Shomaj Company LLC & 86.57.120.0/21 & 40 \\ \hline
PARSONLINE PARSONLINE Autonomous System & 91.98.160.0/19 & 38 \\ \hline
DCI-AS Information Technology Company (ITC) & 217.218.0.0/17 & 37 \\ \hline
IR-ASRETELECOM-AS Asre Enteghal Dadeha & 188.34.48.0/20 & 35 \\ \hline
PARSONLINE PARSONLINE Autonomous System & 91.99.32.0/19 & 34 \\ \hline
IRANGATE Rasaneh Esfahan Net & 212.50.244.0/22 & 33 \\ \hline
DCI-AS Information Technology Company (ITC) & 78.39.0.0/17 & 30 \\ \hline
DCI-AS Information Technology Company (ITC) & 78.39.128.0/17 & 30 \\ \hline
DCI-AS Information Technology Company (ITC) & 85.185.0.0/17 & 30 \\ \hline
ASK-AS Andishe Sabz Khazar Autonomous System & 95.82.40.0/21 & 27 \\ \hline
\hline
\end{tabular}
\caption{Networks Containing Significantly Performant Tests, February 1 2011 - November 29 2011}
\label{appendix:vip:before}
\end{figure}
\begin{figure}
\begin{tabular}{| l | l | c |}
\hline
DCI-AS Information Technology Company (ITC) & 2.176.0.0/16 & 56 \\ \hline
MOBINNET-AS Mobin Net Communication Company & 178.131.0.0/16 & 40 \\ \hline
PARSONLINE PARSONLINE Autonomous System & 91.98.0.0/15 & 36 \\ \hline
PARSONLINE PARSONLINE Autonomous System & 91.99.32.0/19 & 33 \\ \hline
NGSAS Neda Gostar Saba Data Transfer Company & 188.158.0.0/16 & 25 \\ \hline
PARSONLINE PARSONLINE Autonomous System & 91.98.160.0/19 & 25 \\ \hline
PARSONLINE PARSONLINE Autonomous System & 91.99.0.0/19 & 25 \\ \hline
TIC-AS Telecommunication Infrastructure Company & 2.185.128.0/19 & 25 \\ \hline
DCI-AS Information Technology Company (ITC) & 2.181.0.0/16 & 24 \\ \hline
IUSTCC-AS Iran University Of Science and Technology & 194.225.232.0/21 & 24 \\ \hline
TIC-AS Telecommunication Infrastructure Company & 2.180.32.0/19 & 23 \\ \hline
DCI-AS Information Technology Company (ITC) & 46.100.128.0/17 & 22 \\ \hline
PARSONLINE PARSONLINE Autonomous System & 91.99.0.0/16 & 22 \\ \hline
DCI-AS Information Technology Company (ITC) & 2.185.0.0/16 & 21 \\ \hline
NGSAS Neda Gostar Saba Data Transfer Company & 188.158.96.0/21 & 21 \\ \hline
SHARIF-EDU-NET Sharif University of Technology, Tehran,Iran & 213.233.160.0/19 & 21 \\ \hline
DCI-AS Information Technology Company (ITC) & 217.218.0.0/17 & 20 \\ \hline
DCI-AS Information Technology Company (ITC) & 2.182.32.0/19 & 19 \\ \hline
PARSONLINE PARSONLINE Autonomous System & 91.98.208.0/20 & 19 \\ \hline
UT-AS University of Tehran Informatics Center & 80.66.176.0/20 & 19 \\ \hline
FANAVA-AS Fanava Group & 95.38.32.0/19 & 18 \\ \hline
\end{tabular}
\caption{Networks Containing Significantly Performant Tests, November 29 2011 - June 1 2012 }
\label{appendix:vip:during}
\end{figure}


\end{document}